\documentclass[prb,twocolumn,showpacs,preprintnumbers,amsmath,amssymb]{revtex4-1}
\usepackage{graphicx}
\usepackage{dcolumn}
\usepackage{bm}
\usepackage{times}
\usepackage{mathptmx}

\newcounter{saveeqn}

\begin{document}

\title{\emph{Ab-initio} description of heterostructural alloys:
Thermodynamic and structural properties of Mg$_x$Zn$_{1-x}$O and
Cd$_x$Zn$_{1-x}$O}

\author{A. Schleife, M. Eisenacher, C. R{\"o}dl, F. Fuchs, J. Furthm{\"u}ller, and F. Bechstedt}
\affiliation{Institut f\"ur Festk\"orpertheorie und -optik,
Friedrich-Schiller-Universit\"at and European Theoretical
Spectroscopy Facility (ETSF), Max-Wien-Platz 1, 07743 Jena,
 Germany }
\date{\today}

\begin{abstract}
Pseudobinary heterostructural alloys of ZnO with MgO or CdO are
studied by composing the system locally of clusters with varying ratio
of cations. We investigate fourfold (wurtzite structure) and sixfold
(rocksalt structure) coordination of the atoms. By means of 
density functional theory we study a total number of 256 16-atom 
clusters divided into 22 classes for the wurtzite structure and 16
classes for the rocksalt structure for each of the alloy systems. The
fraction with which each cluster contributes to the alloy is
determined for a given temperature $T$ and composition $x$ within (i)
the generalized quasi-chemical approximation, (ii) the model of a
strict-regular solution, and (iii) the model of microscopic
decomposition. From the cluster fractions we derive conclusions about
the miscibility and the critical compositions at which the average
crystal structure changes. Thermodynamic properties such as the mixing
free energy and the mixing entropy are investigated for the three
different statistics. We discuss the consequences of the two different
local lattice structures for characteristic atomic distances, cohesive
energies, and the alloys' elasticities. The differences in the properties
of Mg$_x$Zn$_{1-x}$O and Cd$_x$Zn$_{1-x}$O are explained and
discussed.
\end{abstract}
 \maketitle

\section{Introduction}

Recently, ZnO has attracted renewed attention as a closely
lattice-matched substrate for GaN and also as a potentially useful active
optoelectronic material in its own
right~\cite{Reynolds:1996,Reynolds:1997,Joseph:1999,Ozgur:2004}. When 
doped with Al it plays a role as a transparent conducting oxide in
solar cells~\cite{Special:2000}, but also microelectronic and
sensory devices may be fabricated involving oxides like
ZnO~\cite{Tsukazaki:2007,Ramirez:2007}. Further, it is environmentally
friendly, biocompatible, and tends to form
nanostructures~\cite{SchmidtMende:2007}.

The tailoring of material
properties is an important reason for combining ZnO with 
other group-II oxides in alloys and heterostructures. Mixed
Mg$_x$Zn$_{1-x}$O crystals allow to tune the energy gap from about 3.4~eV of
wurtzite ($wz$) ZnO towards the ultraviolet spectral region with at
least 4.4~eV in
Mg$_x$Zn$_{1-x}$O~\cite{Ohtomo:2005,SchmidtR:2003,Sadofev:2005}. Conversely,
pseudobinary Cd$_x$Zn$_{1-x}$O alloys tend to close the gap 
and are, therefore, suitable candidates for optoelectronic devices in
the visible spectral range~\cite{Sadofev:2006}. Moreover, quantum-well
and other heterostructures based on the combinations
Mg$_x$Zn$_{1-x}$O/ZnO and ZnO/Cd$_x$Zn$_{1-x}$O allow an additional
tailoring of electronic and optical properties by means of
quantum-confinement
effects~\cite{Tsukazaki:2007,Sadofev:2006,Shigemori:2004}.

The combination of these group-II oxides in alloys and
heterostructures raises several fundamental questions, since there is
a crystal structure mismatch: While some binary II-VI compounds
occur in the fourfold-coordinated $wz$ structure (e.g.\ ZnO), others
show the sixfold-coordinated rocksalt ($rs$) structure (e.g.\ MgO,
CdO) under ambient
conditions~\cite{Martienssen:2005,Schleife:2006}. Isovalent and
isostructural alloys of II-VI constituents are generally
thermodynamically unstable, because their mixing enthalpy in either
the $rs$ structure or the $wz$ structure is
positive~\cite{Sanati:2003}. A thermodynamical miscibility only
exists at very high temperatures, i.e., where the mixing entropy is
sufficiently large; at lower temperatures there is a tendency
for a phase separation of the alloys~\cite{Segnit:1965}. However,
isovalent but heterostructural II-VI alloys seem to be stable under
certain conditions: X$_x$Zn$_{1-x}$O alloy films occur in $wz$
structure for $x<0.55$ (X\,=\,Mg)~\cite{Bundesmann:2006,Liu:2009,Du:2009}
or $x<0.32$ (X\,=\,Cd, grown under non-equilibrium
conditions)~\cite{Sadofev:2006,Shigemori:2004}. Depending on
the composition $x$ the film preparation may give rise to a
non-uniformity of pseudobinary thin films as observed experimentally for
Mg$_x$Zn$_{1-x}$O~\cite{Chen:2003}. With increasing molar fractions
$x$ (Mg or Cd content) a change of the coordination (fourfold to
sixfold) of the atoms is expected. Both hexagonal and cubic
Mg$_x$Zn$_{1-x}$O thin films are reported~\cite{Chen:2004}.
The situation in pseudobinary Cd$_x$Zn$_{1-x}$O is less clear. The
$rs$ structure of CdO seems to limit the equilibrium solubility in
$wz$-ZnO to molar fractions below
$x=0.07$~\cite{makino:2001,bertram:2006}. On the other hand, metal-organic
chemical vapor deposition (MOCVD) techniques produced alloys up to
$x=0.7$~\cite{Shigemori:2004}. Molecular beam epitaxy (MBE) leads to
epilayers with a structure close to the $wz$ one without an indication
for a phase separation up to a Cd concentration of
$x=0.32$~\cite{Sadofev:2006}.

Summarizing, there is no microscopic picture of how the
heterostructural aspect influences the properties of isovalent
pseudobinary alloys with oxygen as anion. Stability, solubility, and
tendency of decomposition of systems such as Mg$_x$Zn$_{1-x}$O or 
Cd$_x$Zn$_{1-x}$O are barely understood, so is their different
behavior. One reason is that most of the existing
theoretical studies are limited to ordered structures such as
MgZnO$_2$~\cite{Lambrecht:1999,Thangavel:1999} or somewhat more
complex geometries~\cite{Sanati:2003}.

On the other hand, the theoretical methods for the description of the
thermodynamic and structural properties of pseudobinary alloys
of the type A$_x$B$_{1-x}$C have been developed over the last 20
years. A basic approach is to simulate the configurational average by
representing all possible occurring local bonding configurations
using certain crystal geometries. There
is one study~\cite{Kim:2001} in which the chemically disordered
Mg$_x$Zn$_{1-x}$O alloys have been simulated using special
quasi-random structures~\cite{Wei:1990}. In another
approach~\cite{Chen:1995,Teles:2000} the mixing free energy of a
random alloy is described combining a cluster-expansion
method~\cite{Sanchez:1984,Zunger:1994:SDA,Liu:2008} with the
generalized quasi-chemical approximation (GQCA)~\cite{Sher:1987}. In
several publications about pseudobinary alloys with the first-row
elements N or O as anions~\cite{Kim:2001,Teles:2000,Caetano:2006},
this method has been applied together with an \emph{ab-initio}
description of total energies and atomic geometries.
However, in all these studies the (local) crystal structure has been
fixed being $wz$, $zb$, or $rs$. Only recently the isovalent but
heterostructural MgO-ZnO system has been studied by combining the $rs$
with the $wz$ crystal structure within a cluster expansion
\cite{Sanati:2003,Fan:2008}. Energetic stability has been found in
the sixfold-coordinated structure for Zn concentrations below 67\%,
giving rise to spontaneously ordered
alloys~\cite{Sanati:2003}. Without a configurational average and
studying only certain clusters with fixed molar
fractions~\cite{Fan:2008} no conclusions are possible for different
growth situations. Coherent potential approximation
calculations~\cite{Maznichenko:2009} for various crystal phases 
allowed for the construction of a phase diagram with a phase
transition from $wz$ to $rs$ close to $x=0.33$. However, the coherent
potential approximation, as a
single-site theory, is expected to be more reliable close to the
binary end components, whereas for intermediate compositions $x$ the
single-site approximation is questionable.

In this paper we extend the generalized quasi-chemical approach for one
local crystal structure to alloys that consist of two binary compounds
with two different equilibrium crystal structures, i.e., $rs$ and
$wz$. The results of the GQCA are compared to two other cluster
statistics, that may be considered as limiting cases of the GQCA with
respect to the number of possible atom arrangements. The theory is
applied to heterostructural Mg$_x$Zn$_{1-x}$O and Cd$_x$Zn$_{1-x}$O
alloys. In Section~\ref{sec_th_gqca} we present a cluster
expansion for the $rs$ crystal structure along with the combined
statistics that deals with $wz$ and $rs$ clusters simultaneously.
The computational approaches that are used for carrying out the actual 
calculations are explained in
Section~\ref{sec_compute}. Results for the thermodynamic properties
are given in Section~\ref{sec_thermo} and for the structural and
elastic properties in Section~\ref{sec_lattice}. Finally,
Section~\ref{sec_conclusion} concludes the paper.

\section{\label{sec_th_gqca}Alloy Statistics and Thermodynamics}

\subsection{\label{sec_th_cluster} Cluster expansion for wurtzite and
  rocksalt structures}

We study pseudobinary alloys A$_x$B$_{1-x}$C with $N$ atoms of type C
on the anion sublattice and $N$ atoms of type A or B on the cation
sublattice. Within a cluster expansion
method~\cite{Chen:1995,Teles:2000,Sanchez:1984,Zunger:1994:SDA,Sher:1987}
the macroscopic alloy is divided into an ensemble of $M$ clusters 
consisting of $2n$ atoms ($n$ anions and $n$ cations) each. The total
number of cations or anions is then given by $N=nM$.

From combinatorics it follows that, for a given crystal structure,
there are $2^n$ different possibilities of arranging A- or B-type
atoms on the $n$ cation sites of one cluster (the occupation of the
anion sublattice is fixed). Due to the symmetry of the crystal
lattice, the clusters can be grouped in $J+1$ different classes, with
$J$ depending on the actual crystal structure. Each class $j~(j=0,\dots,J)$ contains $g_j$
clusters of the same total energy $\varepsilon_j$, with the degeneracy
factors $g_j$ fulfilling the relation $\sum_{j}{g_j}=2^n$.

To each macroscopic alloy one can assign a
cluster set $\{M_0, M_1, \dots, M_J\}$ which describes how many
clusters of each class occur in the alloy. A single class $j$
contributes to the macroscopic alloy with its cluster fraction $x_j$
that is defined by $x_j=M_j/M$. The $x_j$ fulfill the constraint
\begin{equation}
\label{eq:cons_1}
\sum^J_{j=0}x_j=1,
\end{equation}
which stems directly from the relation
$M=\sum_j{M_j}$ for the cluster set. The $n$ cation sites of each
cluster are occupied with $n_j$ atoms of species A and $(n-n_j)$ atoms
of species B. Since the molar fraction of A atoms for the entire alloy
A$_x$B$_{1-x}$C is fixed by $x$, the cluster fractions $x_j$ have to
obey the second constraint
\begin{equation}
\label{eq:cons_2}
\sum^J_{j=0}n_jx_j=nx.
\end{equation}
Using such a cluster expansion any macroscopic alloy can be built from
the microscopic clusters, each of which contributes with its cluster
fraction. Consequently, within this framework each property $P$ of the
macroscopic system can be traced back to the respective properties
$P_j$ of the clusters. Given the weights $x_j (x,T)$ for an alloy of a
certain composition $x$ at a temperature $T$ and the values $P_j$ of
the property for each cluster, one can calculate the property $P(x,T)$
for the alloy using the Connolly-Williams
method~\cite{Connolly:1983,Teles:2000},
\begin{equation}
\label{eq:connolly}
P(x,T)=\sum^J_{j=0}x_j(x,T)P_j .
\end{equation}
With this approach structural, elastic, and
thermodynamic properties, including lattice parameters and bulk
moduli can be accessed.

\begin{figure}
\resizebox{0.7\columnwidth}{!}{\includegraphics*{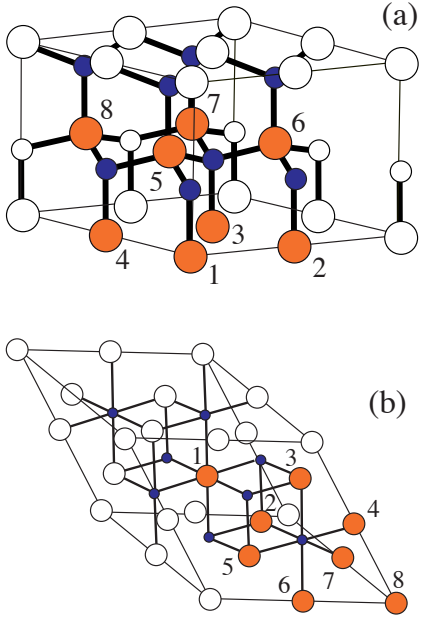}}
\caption{\label{alloys_1}(Color online) Illustration of atomic sites
  in the 16-atom clusters of wurtzite (a) and rocksalt (b)
  structure. Anions (C atoms) are depicted in blue (small), cations
  (A or B atoms) in red (large). White atoms represent the
  surrounding lattice and are not part of the 16-atom clusters. The
  denotation 1\,--\,8 of the cations is used in
  Table~\ref{table_cluster}.}
\end{figure}

Both the Mg$_x$Zn$_{1-x}$O alloy and the Cd$_x$Zn$_{1-x}$O
alloy consist of two monoxides, which show different equilibrium lattice
structures: MgO and CdO crystallize in the cubic $rs$ phase, while the
equilibrium structure of ZnO is the hexagonal $wz$ phase under ambient
conditions. In this work we follow previous
argumentations~\cite{Caetano:2006,Zunger:1990} that it is sufficient
to include next-nearest-neighbor correlations to capture large parts
of the physics involved in the problem. Consequently, we use 16-atom
cluster cells (i.e., $n=8$) and assume that possible correlations
between the different clusters are small.

While we rely on the cluster expansion for the $wz$ structure which is
described in Ref.~\onlinecite{Caetano:2006}, we present a new one
for alloys with $rs$ crystal structure in this
work. The corresponding unit cells and the labeling of the
cation positions for both types, $wz$ and $rs$ clusters, are
illustrated in Fig.~\ref{alloys_1}. In Table~\ref{table_cluster} we
give the number $n_j$ of A cations, the degeneracy factors $g_j$, and
one representative of the symmetry-equivalent clusters for each class
$j$ (characterized by the cation sites that are occupied with A
atoms). The degeneracy $g_j$ for each class as well as the total
number of classes $J+1$ do not only depend on the number of atoms in
the clusters, but also on the actual point group of the crystal. For
$n=8$ we obtain 22 classes in the $wz$ case and 16 
classes for the $rs$ structure (see Table~\ref{table_cluster}). Taking
into account the degeneracies $g_j$ a total of 256 clusters per
crystal structure are studied for each pseudobinary material system.

\begin{table}
\caption{\label{table_cluster}Cluster classes for 16-atom cells in $wz$
  and $rs$ crystal structure. For each class $j$ the number $n_j$ of A
  cations, the degeneracy $g_j$ of the class, and the cation sites
  occupied with A-type atoms (for one representative of the class) are
  given.}
\begin{ruledtabular}
\begin{tabular}{ c c c c|c c c c }
\multicolumn{4}{ c|}{$wz$} & \multicolumn{4}{ c }{$rs$} \\ \hline
 $j$ & $n_j$ & $g_j$ & A atoms & $j$ & $n_j$ & $g_j$ & A atoms \\ \hline
 $0$ & $0$ & $1 $ & --  &$22$ & $0$ & $1$ & -- \\
 $1$ & $1$ & $8 $ & $1  $ & $23$ & $1$ & $8$ & $1$ \\
 $2$ & $2$ & $12$ & $1,2$ & $24$ & $2$ & $24$ & $1,2$ \\
 $3$ & $2$ & $12$ & $1,5$ & $25$ & $2$ & $4$ & $4,5  $ \\
 $4$ & $2$ & $4 $ & $1,8$ & $26$ & $3$ & $32$ & $1,2,3  $ \\
 $5$ & $3$ & $8 $ & $1,2,3$ & $27$ & $3$ & $24$ & $1,4,5  $ \\
 $6$ & $3$ & $24$ & $1,2,7$ & $28$ & $4$ & $8$ & $1,2,3,4 $ \\
 $7$ & $3$ & $24$ & $1,2,5$ & $29$ & $4$ & $8$ & $1,2,3,5 $ \\
 $8$ & $4$ & $2 $ & $1,2,3,4$ & $30$ & $4$ & $48$ & $1,2,4,5  $ \\
 $9$ & $4$ & $8 $ & $1,2,3,5$ & $ 31$ & $4$ & $6$ & $3,4,5,6 $ \\
$10$ & $4$ & $24$ & $1,2,4,5$ & $32$ & $5$ & $32$ & $1,2,3,4,5$ \\
$11$ & $4$ & $6 $ & $1,2,5,6$ & $33$ & $5$ & $24$ & $1,3,4,5,6$ \\
$12$ & $4$ & $6 $ & $1,2,7,8$ & $34$ & $6$ & $24$ & $1,2,3,4,5,6$ \\
$13$ & $4$ & $24$ & $1,2,5,8$ & $35$ & $6$ & $4$ & $2,3,4,5,6,7$ \\
$14$ & $5$ & $24$ & $3,4,6,7,8$ & $36$ & $7$ & $8$ & $1,2,3,4,5,6,7$\\
$15$ & $5$ & $24$ & $3,4,5,6,8$ & $37$ & $8$ & $1$ & $1,2,3,4,5,6,7,8$  \\
$16$ & $5$ & $8 $ & $4,5,6,7,8$ & & & &  \\
$17$ & $6$ & $4 $ & $2,3,4,5,6,7$ & & & & \\
$18$ & $6$ & $12$ & $2,3,4,6,7,8$ & & & & \\
$19$ & $6$ & $12$ & $3,4,5,6,7,8$ & & & \\
$20$ & $7$ & $8$ & $2,3,4,5,6,7,8$ & & & \\
$21$ & $8$ & $1$ & $1,2,3,4,5,6,7,8$ & & & \\
\end{tabular}
\end{ruledtabular}
\end{table}

\subsection{Generalized quasi-chemical approximation}

In the framework of the generalized quasi-chemical approximation
(GQCA),~\cite{Chen:1995,Teles:2000,Sanchez:1984,Zunger:1994:SDA,Sher:1987}
the cluster fractions are determined by a minimization of the
Helmholtz free energy $F(x,T)$. We restrict the discussion to the
mixing contribution $\Delta F$ to the free energy

\begin{equation}
\label{eq:delta_f}
\Delta F(x,T) = \Delta U (x,T) - T \Delta S (x,T).
\end{equation}

The mixing contribution to the internal energy $\Delta U(x,T)$ is
calculated as the sum over the contributions from the $M$ clusters and
is referenced to the value of $U$ of an alloy consisting only of the
two binary end components AC and BC,
\begin{equation}
\label{eq:delta_u}
\begin{split}
\Delta U(x,T)
&=M\left\{\sum^J_{j=0}x_j\varepsilon_j-\left[x\varepsilon_J+(1-x)\varepsilon_0\right]\right\}
\\
&= M\sum_{j=0}^{J}{\Delta\varepsilon_j x_j}.
\end{split}
\end{equation}
Here we introduce the definition of the excess energy
$\Delta\varepsilon_j$ for the class $j$,
\begin{equation}
\label{eq:excess}
\Delta\varepsilon_j = \varepsilon_j - \left(\frac{n_j}{n}\varepsilon_J +
  \frac{n-n_j}{n}\varepsilon_0\right).
\end{equation}

For the calculation of the free energy, Eq.~\eqref{eq:delta_f}, or
thermodynamic properties derived thereof, an expression for the
configurational (or mixing) entropy has to be found.~\cite{Teles:2000}
To evaluate the Boltzmann definition of the entropy, $\Delta S(x,T) =
k_\text{B} \ln W$, one has to 
give an expression for the number of possible configurations
$W$. Given a cluster expansion and, therefore, a set of cluster
fractions $\left\{x_j\right\}$ that fulfills the constraints
\eqref{eq:cons_1} and \eqref{eq:cons_2}, $W$ describes the number of
possible atomic configurations in the entire alloy, i.e., $W$ counts
all possible ways of arranging the $N_\text{A}$ A atoms and $N_\text{B}$ B 
atoms for \emph{one} given set $\left\{x_j\right\}$ on the
$N=N_\text{A}+N_\text{B}$ 
cation sites.
To
determine $W$, the number of ways of arranging the $M_0, M_1, \dots ,
M_J$ clusters to form the alloy, $M!/\prod_{j}M_j!$, has to be
multiplied by the number of possibilities to arrange the cations in each
cluster. Since one cluster of class $j$ can be occupied by cations in
$g_j$ ways, all $M_j$ clusters lead to $g_j^{M_j}$
possibilities. Taking into account all classes $j$ one ultimately
obtains
\begin{equation}
\label{eq:w_exact}
W=\frac{M!}{\prod^J_{j=0}M_j!}\cdot\prod^J_{j'=0}g_{j'}^{M_{j'}}.
\end{equation}

Using this expression for $W$ and the definition $x_j=M_j/M$ one finds
for the mixing entropy $\Delta S$ in the Stirling limit
\begin{equation}
\label{eq:delta_s}
\Delta S(x,T)=-k_\text{B} M \sum^J_{j=0}x_j \ln\left(\frac{x_j}{g_j}\right).
\end{equation}

Introducing the \emph{ideal} cluster fractions of a strict-regular
solution~\cite{Sher:1987} (see Section~\ref{sec:srs_mdm}),
\begin{equation}
\label{eq:x_j^0}
x_j^0 = g_j x^{n_j} \left(1-x\right)^{n-n_j},
\end{equation}
the mixing entropy can be rewritten into~\cite{Sher:1987,Teles:2000}
\begin{equation}
\begin{split}
\label{eq:delta_s_old_papers}
\Delta S(x,T)=-k_\text{B} \Big\{ &N \left[x \ln x + (1-x)\ln (1-x)\right] \\
                      &+ M\sum^J_{j=0}x_j \ln\Big(\frac{x_j}{x_j^0}\Big)\Big\}.
\end{split}
\end{equation}
Although the expressions for $W$ according to Eq.~\eqref{eq:w_exact}
and the one derived in Ref.~\onlinecite{Teles:2000} differ,
Eq.~\eqref{eq:delta_s_old_papers} shows that they both lead
to the same entropy in the Stirling limit.

Equations~\eqref{eq:delta_u} and~\eqref{eq:delta_s} fully determine
the Helmholtz mixing free energy as a function of $x$ and $T$, given
that the cluster fractions $x_j$ are known for $x$ and $T$. In the
GQCA the $x_j$ are determined by the requirement that $\Delta F(x,T)$
assumes a minimum with respect to the cluster distribution, i.e.,
$\partial\Delta F(x,T)/\partial x_j=0$. Hence, the Lagrange formalism
with the constraints~\eqref{eq:cons_1} and~\eqref{eq:cons_2} yields
\begin{equation}
\label{eq:x_j}
x_j^{\rm
  GQCA}(x,T)=\frac{g_j\eta^{n_j}e^{-\beta\Delta\varepsilon_{j}}}{\sum^J_{j'=0}g_{j'}\eta^{n_{j'}}e^{-\beta\Delta\varepsilon_{j'}}},
\end{equation}
whereas $\beta = 1/k_{\rm B} T$. The parameter $\eta$ has to
be determined from the condition that the $x_j^{\rm GQCA}$ obey the
constraint~\eqref{eq:cons_2}. The cluster fractions $x_j^{\rm GQCA}$
according to Eq.~\eqref{eq:x_j} describe the probability for the
occurrence of each cluster class $j$ in an alloy which has been
prepared under thermodynamic equilibrium conditions that minimize the
free energy.

Since we study two cases, isostructural ($wz$ or $rs$ only) as
well as heterostructural (combined $wz$ and $rs$) alloys,
we distinguish three situations: (i) $wz$ clusters only, (ii) $rs$
clusters only, or (iii) simultaneous occurrence of $wz$ and $rs$
clusters. Depending on the clusters included in the calculation, the
value for $J$ has to be set to $J=21$, $15$, or $37$ (see
Table~\ref{table_cluster}) and the minimization 
of $\Delta F$ has to be performed for given $x$ and $T$ accordingly,
which leads, via Eq.~\eqref{eq:x_j}, to the corresponding $x_j^{\rm GQCA}$
for each case.

\subsection{\label{sec:srs_mdm}Strict-regular solution and microscopic decomposition limit}

In the literature several techniques for the deposition of
Mg$_x$Zn$_{1-x}$O films are reported, e.g.\ pulsed laser
deposition (PLD) with growth temperatures of
$950\dots 1050$~K~\cite{SchmidtR:2003}, radio frequency magnetron
sputtering at $700$~K~\cite{Jeong:2003}, and reactive 
electron beam evaporation (REBE) at a substrate temperature of
$550$~K~\cite{Chen:2003}. Similar techniques have been applied for
Cd$_x$Zn$_{1-x}$O layers, however, at much lower substrate temperatures,
e.g.\ MBE with a growth temperature as low as
$450$~K~\cite{Sadofev:2006}, (plasma-enhanced) MOCVD at a growth
temperature of $625$~K~\cite{Shigemori:2004,bertram:2006}, or PLD at
$700$~K~\cite{makino:2001}. Subsequent repeated temper steps are
reported in some cases. The films are typically deposited on $c$-plane (0001)
sapphire, but also $a$-plane sapphire or more exotic substrates have been
used. Therefore, the respective experimental situation suggests the
studying of certain non-equilibrium preparation conditions, for which
the actual cluster statistics may be modified by kinetic barriers, frozen 
high-temperature states, as well as interface or surface
influences. In order to simulate a dependence of the cluster
distribution on the preparation conditions we study two limiting
cases:

(i) The strict-regular solution (SRS) model~\cite{Sher:1987}. In this
case, the \emph{ideal} cluster fractions $x_j^0$ according to
Eq.~\eqref{eq:x_j^0} are used which arise from a purely stochastic
distribution of the clusters. These $x_j^0$ do neither depend on the
temperature nor the clusters' excess energies but are only determined by
$x$ and $n_j$. The number $W$ is then
simply given by all possible arrangements of $N_\text{A}=xN$ A atoms
and $N_\text{B}=(1-x)N$ B atoms on the $N=N_\text{A}+N_\text{B}$
cation sites of the alloy, i.e.,
\begin{equation}
\label{eq:w_ideal}
W^{\rm SRS}=\frac{N!}{N_\text{A}!N_\text{B}!}.
\end{equation}
In the Stirling limit the mixing entropy in this case reduces to
\begin{equation}
\label{eq:delta_s_ideal}
\Delta S^{\rm SRS}(x)=-k_\text{B} N \left[x \ln x + (1-x)\ln (1-x)\right].
\end{equation}
The ideal $x_j^0$ according to Eq.~\eqref{eq:x_j^0} and $\Delta S^{\rm
SRS}$ can be interpreted as the high-temperature limit of the GQCA, since 
for increasing temperatures the $x_j^{\rm GQCA}$ approach the $x_j^0$ and,
consequently, $\Delta S$ [cf.\ Eq.~\eqref{eq:delta_s}] approaches
$\Delta S^{\rm SRS}$.

(ii) The microscopic decomposition model (MDM). In this limiting case
the cations of type A (B) are more likely to occur close to
cations of type A (B). Consequently, only the clusters representing the
two binary components AC and BC are allowed, with $xM$ being the
number of AC clusters and $(1-x)M$ the number of BC clusters. This is
equivalent to merely a linear interpolation between the 
binary end components and, therefore, the results are equal to
predictions from Vegard's
rule~\cite{Vegard:1921,Teles:2000}. Furthermore, for 
positive excess energies $\Delta\varepsilon_j$
[cf.\ Eq.~\eqref{eq:excess}] the $x_j^{\rm MDM}$ represent the
low-temperature limit of the GQCA. The cluster fractions for the MDM
are given by
\begin{equation}
\label{eq:x_j^mdm}
x_j^{\rm MDM} =
\left\{
\begin{array}{cc}
1-x & {\rm for}~j=0 \\
x   & {\rm for}~j=J \\
0   & {\rm otherwise}
\end{array}
\right. .
\end{equation}
The number of atomic configurations follows immediately 
from~Eq.~\eqref{eq:w_exact} by taking into account only the two
clusters with $j=0$ and $j=J$, i.e., 
\begin{equation}
\label{eq:w_mdm}
W^{\rm MDM}=\frac{M!}{\left[xM\right]!\left[(1-x)M\right]!}.
\end{equation}
Using the $x_j^{\rm MDM}$ we obtain for the mixing entropy
\begin{equation}
\begin{split}
\label{eq:delta_s_mdm}
\Delta S^{\rm MDM} (x) &= -k_{\rm B} M \left[x \ln x + (1-x)\ln (1-x)\right] \\
                       &= \frac{1}{n} \Delta S^{\rm SRS} (x),
\end{split}
\end{equation}
i.e., a reduced configurational entropy. In the MDM the resulting
length scale of decomposition of the alloy into the binary compounds
is not well defined as usual in a thermodynamic approach. The minimum
length scale is given by the assumed cluster size itself. Whether
clusters of the same type (AC or BC) build larger regions of pure AC
and BC material on a mesoscopic or macroscopic length scale depends on
the cluster interaction (which is not taken into account). In general,
the MDM describes alloys that have been prepared under conditions
where mixing does not lead to a gain of internal energy. Indeed, in
the MDM description one finds from Eq.~\eqref{eq:delta_u} $\Delta
U(x)=0$ for the mixing contribution to the internal energy.

\section{\label{sec_compute}Quantum Mechanical Calculations of Clusters}

\subsection{Computational methods}

The statistical methods introduced in Section~\ref{sec_th_gqca}
trace the properties of an alloy of composition $x$ at temperature
$T$ back to the corresponding properties of individual clusters using
Eq.~\eqref{eq:connolly}. Therefore, the respective calculations have
to be performed for each of the 22 ($wz$) and 16 ($rs$) clusters of
Mg$_x$Zn$_{1-x}$O and Cd$_x$Zn$_{1-x}$O.

Ground-state properties, such as the total energy of a cluster or its
structural properties, are well described within density functional
theory (DFT) \cite{Hohenberg:1964:IEG,Kohn:1965:SCE} applying the
supercell method, where a cluster is identified with the unit cell of
an artificial crystal. Explicit calculations are performed using the
Vienna \emph{Ab-initio} Simulation Package
(VASP)~\cite{Kresse:1996:PRB}. For the semi-local generalized-gradient
approximation (GGA) to the exchange and correlation (XC) potential in
the Kohn-Sham equation~\cite{Kohn:1965:SCE} we rely on the PW91
parametrization as suggested by Perdew and Wang
\cite{Perdew:1991:ESS,Perdew:wang:1992}. Although such a semi-local
treatment of XC cannot correctly describe excitation properties (such
as QP energies or excitonic
effects~\cite{Schleife:2009pss,Schleife:2009:PRB}), it allows for a
very good description of
lattice constants and cohesive energies. For the oxides studied in this work,
the GGA tends to slightly underestimate binding energies and slightly
overestimate lattice constants \cite{Schleife:2006}. However, these
ground-state properties are not systematically improved by a different
treatment of XC, for instance by adding an additional on-site Hubbard
interaction $U$~\cite{Janotti:2006} or a certain amount of Fock
exchange~\cite{Oba:2008}.

The electron-ion interaction is modeled within the projector-augmented
wave (PAW) method \cite{Blochl:1994:PAW,Kresse:1999}. We treat the
Mg\,$2p$ and Mg\,$3s$ as well as the O\,$2s$ and O\,$2p$ electrons as
valence electrons. For Zn and Cd, the Zn\,$3d$ and Cd\,$4d$ states
have to be included as valence states, respectively, together with the
Zn\,$4s$ and Cd\,$5s$ electrons. The electronic wave functions are
expanded into plane waves up to a cutoff energy of $450$~eV and the
Brillouin zone (BZ) is sampled by 2$\times$2$\times$2 ($wz$) or
4$\times$4$\times$4 ($rs$) Monkhorst-Pack (MP) ${\bf k}$
points\cite{Monkhorst:1976}. We checked that these parameters give
results for the total energies of the clusters that are converged
within 0.08\,\%. This corresponds to a very small level of inaccuracy,
especially since the total energies of the different binary end
components differ by at least 10\,\% of their total energy.

In order to obtain the equilibrium values for the cell volumes, total
energies, and bulk 
moduli we minimize the total energy for several cell volumes and fit
the resulting volume dependence of the total energy to the Murnaghan
equation of state\cite{Murnaghan:1944}. For each cluster geometry, we
compute fully relaxed atomic positions, i.e., optimized lattice
constants and internal cell parameters, ensuring that the forces
acting on the ions are well below 5~meV/{\AA}.

\subsection{Results}

To describe the geometry of the two different crystal structures on the
same footing we discuss the average cation-cation (second-nearest
neighbor) distance $d_2$. The 
results for the cluster energies $\varepsilon_j$, the cation-cation
distances $d_{2,j}$ (averaged over all cation-cation distances of one
cluster), the $wz$ lattice parameters $c_j$ of the cluster cell and
$u_j$ (averaged over all $u$'s of one cluster), the 
cell volumes $V_j$, and the bulk moduli $B_{0,j}$ are listed in
Table~\ref{tab:params_wz} for all $wz$ cluster cells and
Table~\ref{tab:params_rs} for all cluster classes of the $rs$ crystal
structure.

The energies of the clusters $\varepsilon_j$ in Tables~\ref{tab:params_wz}
and~\ref{tab:params_rs} show an almost linear variation with the
number $n_j$ of Mg (Cd) atoms, decreasing (increasing) from pure ZnO
to pure MgO (CdO). Small negative (positive) non-linear deviations appear for
Mg$_{n_j}$Zn$_{8-n_j}$O$_8$ (Cd$_{n_j}$Zn$_{8-n_j}$O$_8$) with absolute values
of a few meV (tenths of an eV).

The cation-cation distances $d_{2,j}$ show a weak decrease
(strong increase) with the number $n_j$ of Mg (Cd) atoms as expected
from the corresponding values for the binary end components. While the 
nearest-neighbor distances or bond lengths (not listed in
Tables~\ref{tab:params_wz} and~\ref{tab:params_rs}) of the ZnO-MgO
system decrease only very slightly from $d_{\text{Zn-O}}\approx
2.002/2.167$~\AA\ to~$d_{\text{Mg-O}}\approx 1.997/2.128~$\AA\ for
$wz$/$rs$, we find a completely different situation for the ZnO-CdO
system, where the bond lengths $d_{\text{Cd-O}}\approx
2.230/2.388$~\AA\ are rather different from $d_{\text{Zn-O}}$. For the
bulk moduli $B_{0,j}$ we observe a decrease with increasing number
$n_j$ of Mg or Cd atoms, independent of the crystal 
structure. This trend follows the behavior of the covalent radii of
the cations that also show an anomalous trend of 1.36~\AA, 1.25~\AA,
and 1.48~\AA~along the row Mg, Zn, and Cd~\cite{sargent}. For our
calculated lattice parameters, equilibrium volumes, and bond lengths
we find very good agreement with a similar
study~\cite{Fan:2008}. However, the deviations are remarkable for the 
total energies and, consequently, lead to different results for, e.g.,
the mixing free energy. We attribute these differences to the use of
the local-density approximation (LDA) to XC, and to the neglect of the
Zn\,$3d$ and Cd\,$4d$ electrons as valence electrons in
Ref.~\onlinecite{Fan:2008}.

\begin{table}
\caption{\label{tab:params_wz}Cluster energies per cation-anion pair
  $\varepsilon_j$, cation-cation distances $d_{2,j}$, lattice parameters $c_j$ and $u_j$,
  volume per cation-anion pair $V_j$, and bulk moduli $B_{0,j}$ for 16-atom
    clusters of Mg$_{n_j}$Zn$_{8-n_j}$O$_8$ (first line of each class)
    and Cd$_{n_j}$Zn$_{8-n_j}$O$_8$ (second line of each class) in 
    $wz$ structure.}
\begin{ruledtabular}
\begin{tabular}{ c c c c c c c }
class $j$ & $\varepsilon_j$ & $d_{2,j}$ & $c_j$& $u_j$ & $V_j$ & $B_{0,j}$\\ 
&(eV/pair) & (\AA) &(\AA)&   &(\AA$^3$) & (GPa) \\ \hline
$0$  & $-9.05$ & $3.270$ & $5.303$ & $0.378$ & $24.72$ & $141.0$ \\ \hline
$1$  & $-9.41$ & $3.268$ & $5.288$ & $0.380$ & $24.68$ & $137.5$ \\ 
    & $-8.87$ & $3.315$ & $5.368$ & $0.379$ & $25.75$ & $121.3$ \\ \hline 
$2$  & $-9.76$ & $3.270$ & $5.275$ & $0.381$ & $24.73$ & $131.6$ \\ 
    & $-8.69$ & $3.359$ & $5.452$ & $0.378$ & $26.80$ & $110.2$ \\ \hline 
$3$  & $-9.76$ & $3.266$ & $5.275$ & $0.381$ & $24.64$ & $131.3$ \\ 
    & $-8.70$ & $3.357$ & $5.405$ & $0.382$ & $26.74$ & $112.7$ \\ \hline 
$4$  & $-9.76$ & $3.266$ & $5.271$ & $0.381$ & $24.64$ & $131.1$ \\ 
    & $-8.71$ & $3.358$ & $5.445$ & $0.379$ & $26.77$ & $112.5$ \\ \hline 
$5$  & $-10.12$ & $3.265$ & $5.243$ & $0.383$ & $24.61$ & $121.5$ \\ 
    & $-8.51$ & $3.408$ & $5.565$ & $0.374$ & $27.98$ & $114.2$ \\ \hline 
$6$  & $-10.12$ & $3.265$ & $5.252$ & $0.382$ & $24.60$ & $121.6$ \\ 
    & $-8.55$ & $3.405$ & $5.508$ & $0.380$ & $27.91$ & $117.2$ \\ \hline 
$7$  & $-10.12$ & $3.265$ & $5.260$ & $0.382$ & $24.60$ & $121.4$ \\ 
    & $-8.54$ & $3.404$ & $5.459$ & $0.384$ & $27.87$ & $116.0$ \\ \hline 
$8$  & $-10.47$ & $3.264$ & $5.211$ & $0.385$ & $24.57$ & $118.4$ \\ 
    & $-8.33$ & $3.455$ & $5.705$ & $0.368$ & $29.16$ & $103.7$ \\ \hline 
$9$  & $-10.47$ & $3.264$ & $5.235$ & $0.384$ & $24.58$ & $118.1$ \\ 
     & $-8.37$ & $3.451$ & $5.542$ & $0.383$ & $29.03$ & $104.0$ \\ \hline 
$10$ & $-10.47$ & $3.264$ & $5.232$ & $0.384$ & $24.57$ & $118.0$ \\ 
     & $-8.38$ & $3.452$ & $5.591$ & $0.378$ & $29.07$ & $105.9$ \\ \hline 
$11$ & $-10.47$ & $3.264$ & $5.246$ & $0.383$ & $24.58$ & $117.3$ \\ 
     & $-8.39$ & $3.447$ & $5.467$ & $0.390$ & $28.89$ & $104.1$ \\ \hline 
$12$ & $-10.47$ & $3.263$ & $5.233$ & $0.384$ & $24.57$ & $118.8$ \\ 
     & $-8.41$ & $3.450$ & $5.583$ & $0.380$ & $29.03$ & $107.6$ \\ \hline 
$13$ & $-10.47$ & $3.264$ & $5.241$ & $0.383$ & $24.57$ & $117.6$ \\ 
     & $-8.40$ & $3.449$ & $5.531$ & $0.384$ & $28.98$ & $105.1$ \\ \hline 
$14$ & $-10.82$ & $3.261$ & $5.216$ & $0.385$ & $24.52$ & $116.8$ \\ 
     & $-8.24$ & $3.495$ & $5.576$ & $0.387$ & $30.13$ & $95.9$ \\ \hline 
$15$ & $-10.82$ & $3.261$ & $5.210$ & $0.385$ & $24.51$ & $116.7$ \\ 
     & $-8.25$ & $3.497$ & $5.636$ & $0.381$ & $30.22$ & $98.2$ \\ \hline 
$16$ & $-10.82$ & $3.262$ & $5.198$ & $0.386$ & $24.53$ & $114.7$ \\ 
     & $-8.21$ & $3.501$ & $5.711$ & $0.374$ & $30.33$ & $96.3$ \\ \hline 
$17$ & $-11.17$ & $3.260$ & $5.183$ & $0.387$ & $24.48$ & $113.2$ \\ 
     & $-8.11$ & $3.544$ & $5.714$ & $0.382$ & $31.48$ & $98.4$ \\ \hline 
$18$ & $-11.17$ & $3.260$ & $5.191$ & $0.387$ & $24.49$ & $110.3$ \\ 
     & $-8.10$ & $3.542$ & $5.647$ & $0.387$ & $31.38$ & $96.6$ \\ \hline 
$19$ & $-11.17$ & $3.261$ & $5.182$ & $0.387$ & $24.50$ & $109.8$ \\ 
     & $-8.08$ & $3.546$ & $5.727$ & $0.380$ & $31.51$ & $96.0$ \\ \hline 
$20$ & $-11.52$ & $3.259$ & $5.158$ & $0.389$ & $24.46$ & $110.5$ \\ 
     & $-7.95$ & $3.592$ & $5.762$ & $0.384$ & $32.74$ & $92.4$ \\ \hline 
$21$ & $-11.86$ & $3.258$ & $5.120$ & $0.391$ & $24.41$ & $104.4$ \\ 
     & $-7.82$ & $3.640$ & $5.841$ & $0.384$ & $34.10$ & $90.6$ \\
\end{tabular}
\end{ruledtabular}
\end{table}

\begin{table}
\caption{\label{tab:params_rs}Cluster energies per cation-anion pair
  $\varepsilon_j$, cation-cation distances $d_{2,j}$, volume per
  cation-anion pair $V_j$, and bulk moduli $B_{0,j}$
  for 16-atom clusters of Mg$_{n_j}$Zn$_{8-n_j}$O$_8$ (first line of each class) and
  Cd$_{n_j}$Zn$_{8-n_j}$O$_8$ (second line of each class) in $rs$ structure.}
\begin{ruledtabular}
\begin{tabular}{ c c c c c }
class $j$ & $\varepsilon_j$& $d_{2,j}$& $V_j$ & $B_{0,j}$ \\
& (eV/pair) & (\AA) &(\AA$^3$) & (GPa) \\ \hline
$0$ & $-8.75$ & $3.065$ & $20.36$ & $166.3$ \\ \hline
$1$ & $-9.16$ & $3.059$ & $20.23$ & $165.4$ \\ 
  & $-8.57$ & $3.111$ & $21.29$ & $162.1$ \\ \hline 
$2$ & $-9.57$ & $3.052$ & $20.09$ & $162.7$ \\ 
  & $-8.42$ & $3.152$ & $22.14$ & $154.9$ \\ \hline 
$3$ & $-9.57$ & $3.053$ & $20.11$ & $162.9$ \\ 
   & $-8.34$ & $3.165$ & $22.43$ & $155.4$ \\ \hline 
$4$ & $-9.98$ & $3.044$ & $19.95$ & $160.9$ \\ 
   & $-8.31$ & $3.189$ & $22.92$ & $147.5$ \\ \hline 
$5$ & $-9.98$ & $3.045$ & $19.97$ & $155.0$ \\ 
   & $-8.24$ & $3.202$ & $23.21$ & $147.3$ \\ \hline 
$6$ & $-10.39$ & $3.037$ & $19.81$ & $158.5$ \\ 
   & $-8.22$ & $3.224$ & $23.69$ & $144.9$ \\ \hline 
$7$ & $-10.39$ & $3.037$ & $19.80$ & $158.7$ \\ 
   & $-8.22$ & $3.224$ & $23.70$ & $144.9$ \\ \hline 
$8$ & $-10.39$ & $3.038$ & $19.83$ & $159.0$ \\ 
   & $-8.15$ & $3.236$ & $23.95$ & $145.0$ \\ \hline 
$9$ & $-10.39$ & $3.039$ & $19.85$ & $159.2$ \\ 
  & $-8.09$ & $3.247$ & $24.22$ & $144.3$ \\ \hline 
$10$ & $-10.80$ & $3.030$ & $19.67$ & $156.7$ \\ 
   & $-8.09$ & $3.267$ & $24.66$ & $138.4$ \\ \hline 
$11$ & $-10.80$ & $3.031$ & $19.70$ & $139.5$ \\ 
   & $-8.03$ & $3.280$ & $24.95$ & $132.5$ \\ \hline 
$12$ & $-11.20$ & $3.023$ & $19.54$ & $156.3$ \\ 
   & $-7.99$ & $3.315$ & $25.75$ & $114.1$ \\ \hline 
$13$ & $-11.21$ & $3.024$ & $19.56$ & $157.4$ \\ 
   & $-8.00$ & $3.344$ & $26.39$ & $101.8$ \\ \hline 
$14$ & $-11.61$ & $3.016$ & $19.40$ & $153.5$ \\ 
   & $-7.90$ & $3.343$ & $26.42$ & $130.8$ \\ \hline 
$15$ & $-12.01$ & $3.009$ & $19.26$ & $152.0$ \\ 
   & $-7.84$ & $3.378$ & $27.25$ & $126.9$ \\
\end{tabular}
\end{ruledtabular}
\end{table}

\section{\label{sec_thermo}Thermodynamic properties}

By including the temperature-dependent mixing entropy we also obtain a
temperature dependence of the cluster fractions and, therefore, of the
derived properties. In this work we discuss the alloy properties at
two selected temperatures, (i) room
temperature ($T=300$~K) where many measurements are performed,
and (ii) an exemplary growth temperature of $T=1100$~K.

\subsection{Excess energies}

\begin{figure*}
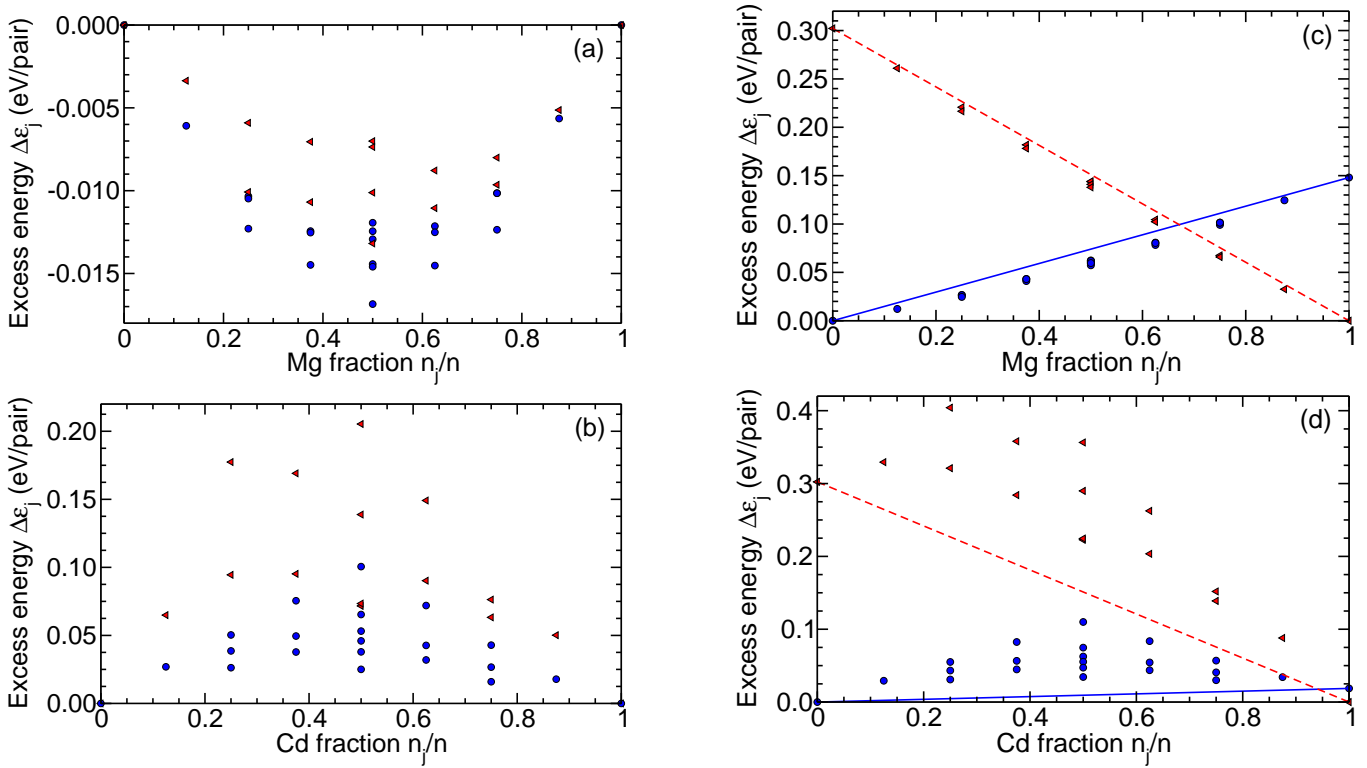

\resizebox{0.46\textwidth}{!}{\includegraphics*{./excess_mgzno.eps}}\hfill\resizebox{0.46\textwidth}{!}{\includegraphics*{./excess_mgzno_b.eps}}

\resizebox{0.46\textwidth}{!}{\includegraphics*{./excess_cdzno.eps}}\hfill\resizebox{0.46\textwidth}{!}{\includegraphics*{./excess_cdzno_b.eps}}
\caption{\label{fig:excess}(Color online) Excess energies per
  cation-anion pair
  $\Delta\varepsilon_j$ of the Mg$_{n_j}$Zn$_{n-n_j}$O$_n$ (a), (c) and
  Cd$_{n_j}$Zn$_{n-n_j}$O$_n$ (b), (d) clusters computed from
  Eq.~(\ref{eq:excess}). The
  excess energies are shown for the rocksalt (red triangles) and the
  wurtzite (blue circles) crystal structures. In subfigures (a) and (b) [subfigures (c) and (d)] the $\varepsilon_0$ and $\varepsilon_J$ of the
  same crystal structure [equilibrium crystal structure] have been
  used as level of reference. The lines linearly connect the end
  points $wz$-ZnO and $wz$-XO (solid blue) as well as $rs$-ZnO and $rs$-XO
  (dashed red), X\,=\,Mg,Cd.}
\end{figure*}

The excess energy [cf.\ Eq.~\eqref{eq:excess}] $\Delta\varepsilon_j$ 
represents the energy of formation of a certain
X$_{n_{j}}$Zn$_{n-n_{j}}$O$_n$ cluster in the $wz$ or $rs$ structure
with respect to ZnO and XO (X\,=\,Mg, Cd) in the respective crystal
structure. They are fundamental quantities for the GQCA, since they ultimately
determine the cluster fractions [see Eq.~\eqref{eq:x_j}] and, therefore,
the alloy statistics. In Fig.~\ref{fig:excess} the excess energies are
plotted for both material combinations, i.e.\ ZnO-MgO and ZnO-CdO. Whereas in 
Figs.~\ref{fig:excess}(a) and~(b) the $\Delta\epsilon_j$ for $wz$ and
$rs$ structure are shown, we visualize the energetic difference
of these two lattice structures by plotting the $\Delta\varepsilon_j$
using the value of $\varepsilon_0$ for $wz$-ZnO and $\varepsilon_J$
for $rs$-XO in Figs.~\ref{fig:excess}(c) and (d).

For MgZnO clusters [cf.\ Fig.~\ref{fig:excess}(a)] we find negative
excess energies (per anion-cation pair) with absolute values of less than
20~meV for both crystal structures. Contrary, for CdZnO clusters
[cf.\ Fig.~\ref{fig:excess}(b)] 
the excess energies are positive for both polymorphs and by one order
of magnitude larger than the ones for MgZnO. As a consequence, in the
isostructural limit with \emph{only} a fourfold ($wz$) or a sixfold
($rs$) coordinated bonding configuration, we predict that mixing is
possible for the ZnO-MgO system independent of the composition $x$,
temperature $T$, and the boundary condition that
fixes the cluster fractions. In the case of the ZnO-CdO system with
its large difference of the bond lengths
$d_{\text{Zn-O}}$\,/\,$d_{\text{Cd-O}}$ and, hence, a remarkable
strain of a bond in the environment of the other cation species, the
situation is different and will be discussed in detail in
Section~\ref{sec:mixing_free_energy}.

Due to the changed reference energy of the excess energies in
Fig.~\ref{fig:excess}(c) and their small absolute values in the case of
ZnO-MgO, we find an almost linear variation between the end points. The
crossing point near $n_j/n\approx 2/3$ indicates that for small Mg 
fractions the fourfold coordination is more stable while for larger Mg
fractions the sixfold one seems to be energetically favored. This is
in agreement with the occurrence of the cubic crystal structure above a
composition of $x\approx 0.67$ in PLD
samples~\cite{Bundesmann:2006}. However, there are also
experimental~\cite{Chen:2003} or other theoretical
results~\cite{Maznichenko:2009,Sanati:2003,Fan:2008} contradicting our
findings. More precisely, Refs.~\onlinecite{Sanati:2003,Fan:2008} find
for $rs$-ZnO and $wz$-ZnO almost the same energy difference as we do
for $rs$-MgO and $wz$-MgO and vice versa. There are strong indications
that the use of the LDA instead of the GGA can cause such deviations
of the total energy differences. Our presumption is supported by a
comparison to other GGA calculations that confirm our result of 0.3~eV
as the energy difference between $rs$-ZnO and $wz$-ZnO as well as the
0.15~eV between $rs$-MgO and
$wz$-MgO~\cite{Zhang:2007:APL,Carrasco:2007}. LDA values that have
been reported seem to lead to opposite differences for MgO and
ZnO~\cite{Fan:2008,Sanati:2003,Seko:2005,Limpijumnong:2004,Maznichenko:2009}.
In addition, Fan~\emph{et al.}\cite{Fan:2008} do not include the
$d$ electrons of Zn or Cd in their calculations which may further influence
the $rs$-$wz$ splittings.

We observe a different situation for the ZnO-CdO system
[cf.\ Fig.~\ref{fig:excess}(d)], since the absolute
values of the excess energies are comparable to the energy differences
of the $rs$ and $wz$ polymorphs. Consequently, the
deviations from the linear interpolation between the end components
are much larger. Furthermore, the $wz$ and $rs$ polymorphs of CdO are
energetically much closer to each other than the corresponding ZnO
polymorphs. Merely plotting the excess energies [see
Fig.~\ref{fig:excess}(d)] suggests a wide range up to $n_j/n\approx
0.95$ for the stability of $wz$ which contradicts the majority of
experiments~\cite{makino:2001,bertram:2006,Sadofev:2006,Shigemori:2004,Ishihara:2006}. Taking
the cluster statistics into account we discuss this issue in more detail later
[cf.\ Sections~\ref{sec:mixing_free_energy} and~\ref{sec:phase_transition}].

\subsection{Internal energy and entropy}

\begin{figure}
\resizebox{0.92\columnwidth}{!}{\includegraphics*{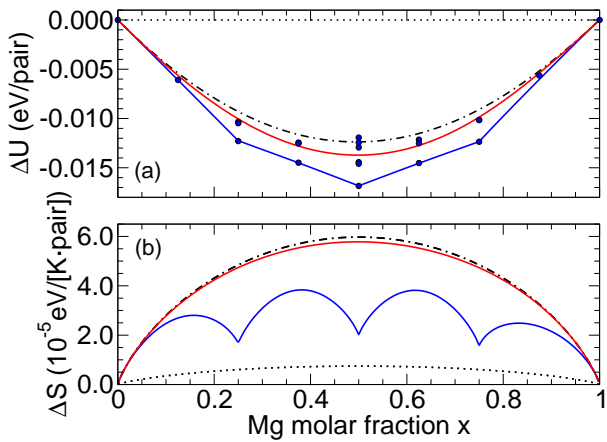}}
\caption{\label{fig:duds}(Color online) Mixing contribution to the
  internal energy $\Delta U(x,T)$ (a) and mixing entropy $\Delta
  S(x,T)$ (b) for Mg$_x$Zn$_{1-x}$O in $wz$ structure. The solid curves are
  obtained using the GQCA cluster fractions for $T=30$~K (blue)
  and $T=300$~K (red). The dotted curves result from the MDM and the
  dash-dotted curves are obtained using the ideal cluster
  fractions. For comparison the blue dots represent the excess
  energies computed from Eq.~(\ref{eq:excess}). All quantities are
  normalized to cation-anion pairs.}
\end{figure}

For a general discussion of the thermodynamic trends resulting from
the three different statistics we compare the contributions $\Delta U$
[Eq.~\eqref{eq:delta_u}] and $\Delta S$
[Eq.~\eqref{eq:delta_s_old_papers}] to the mixing free energy $\Delta
F$ as calculated from the GQCA with
results obtained using the ideal weights, and the weights from the MDM
for $wz$-Mg$_x$Zn$_{1-x}$O.

From the mixing contribution to the internal energy $\Delta U(x,T)$ of
$wz$-Mg$_x$Zn$_{1-x}$O it can be seen [cf.\ Fig.~\ref{fig:duds}(a)] that
for low temperatures ($T=30$~K) only the clusters with the lowest energies
contribute in the GQCA. For higher temperatures ($T=300$~K) the
weights of clusters with higher excess energies are larger. The
SRS model finally represents the high-temperature limit of
the GQCA in which all $wz$ clusters contribute. The MDM does not lower
the internal energy of the system.

In Fig.~\ref{fig:duds}(b) the corresponding results for the mixing
entropy $\Delta S(x,T)$ are shown. Again, the MDM is the most trivial
case with the smallest entropy term, since only two clusters
contribute. The SRS model is obviously maximizing the entropy 
which is consistent with what we expect from a high-temperature
limit, i.e., the high-temperature curve of the GQCA approaches the
SRS limit. The low-temperature curve of the
GQCA shows an interesting behavior, since the entropy in this
case clearly features pronounced minima for compositions $x$ where only
single clusters substantially contribute.

\subsection{\label{sec:mixing_free_energy}Mixing free energy}

\begin{figure}
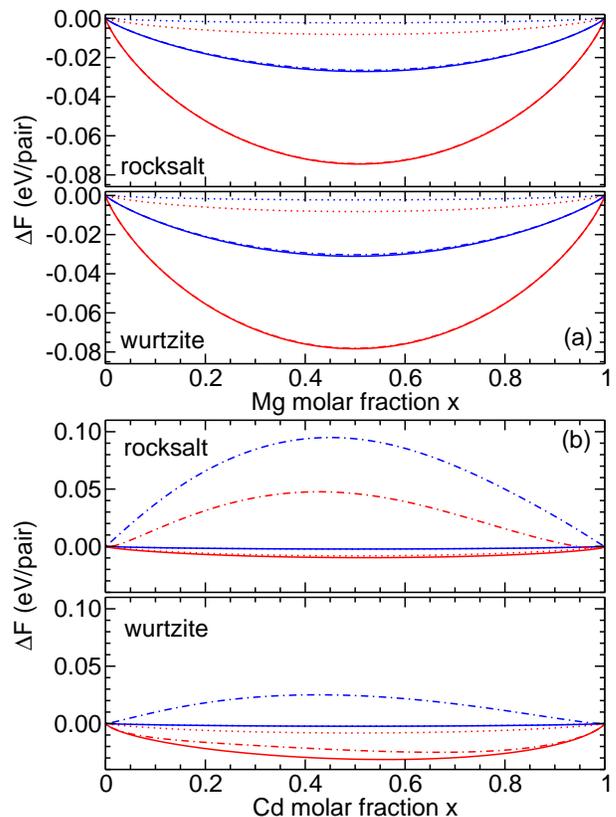

\resizebox{0.92\columnwidth}{!}{\includegraphics*{./dF_rs_wz_MgZnO.eps}}
\resizebox{0.92\columnwidth}{!}{\includegraphics*{./dF_rs_wz_CdZnO.eps}}
\caption{\label{fig:mixing}(Color online) Mixing free energy $\Delta F(x,T)$ of
Mg$_x$Zn$_{1-x}$O (a) and Cd$_x$Zn$_{1-x}$O (b) alloys with fixed
sixfold ($rs$) or fourfold ($wz$) bonding configuration versus
composition $x$. The solid curves are obtained using cluster fractions
from the GQCA. The dotted curves 
result from the MDM and the dash-dotted curves are obtained using the
ideal cluster fractions. In each case we depict results for
$T=300$~K (blue curves) and $T=1100$~K (red curves).}
\end{figure}

The Gibbs free energy is the thermodynamic potential which describes
the equilibrium of a system for fixed temperature and pressure. At low
pressures of about 1 atmosphere the difference between the Gibbs free
energy and the Helmholtz free energy vanishes. Therefore, the
Helmholtz mixing free energy is used as the central thermodynamic
quantity that governs, at least for solids and low pressures, the equilibrium
alloy state and, therefore, the cluster fractions $x_j^{\rm
  GQCA}$.

In Fig.~\ref{fig:mixing} the mixing free energy $\Delta F(x,T)$ for
Mg$_x$Zn$_{1-x}$O and Cd$_x$Zn$_{1-x}$O alloys in the $wz$ and
the $rs$ crystal structures is plotted versus composition $x$. For
Mg$_x$Zn$_{1-x}$O it is obvious that $\Delta F < 0$ for all $x$ and
$T$, which is, according to Eqs.~\eqref{eq:delta_f}
and~\eqref{eq:delta_u}, a direct consequence of the negative excess
energies for Mg$_{n_j}$Zn$_{n-n_j}$O$_n$ clusters
[cf.\ Fig.~\ref{fig:excess}(a)]. For that reason all three statistics
agree in predicting $wz$- and $rs$-Mg$_x$Zn$_{1-x}$O to be a random
alloy without a miscibility gap~\cite{Sher:1987}, independent of $x$
and $T$. There is no tendency for binodal or spinodal decomposition of
the alloy. The results using the GQCA cluster weights and the ones
from the SRS model are almost identical, while the energy gain is
smaller within the MDM.

For Cd$_x$Zn$_{1-x}$O the behavior of $\Delta F$ is qualitatively
different as can be seen from Fig.~\ref{fig:mixing}(b). It shows
that for $T=300$~K and both crystal structures the GQCA results agree
very well with the MDM, i.e., we find the alloy being almost entirely
decomposed into the binary clusters. For the higher temperature of
$T=1100$~K the GQCA result approaches the SRS curve for the
wurtzitic alloy, especially for $x<0.1$ and $x>0.8$. However, for 
$rs$-Cd$_x$Zn$_{1-x}$O at $T=1100$~K the GQCA curve still agrees with the MDM. In
this case the system has a strong tendency for decomposition into the
binary materials and the SRS limit can only be reached at very
high temperatures. Besides, for Cd$_x$Zn$_{1-x}$O the
$\Delta F$ curves versus $x$ are rather asymmetric for the SRS and show
minima and inflection points with positions that strongly depend on
the temperature and the local crystal structure. This is an indication of a
phase transition between random and phase-separated alloys and will be
discussed as such later.

\begin{figure}
\resizebox{0.92\columnwidth}{!}{\includegraphics*{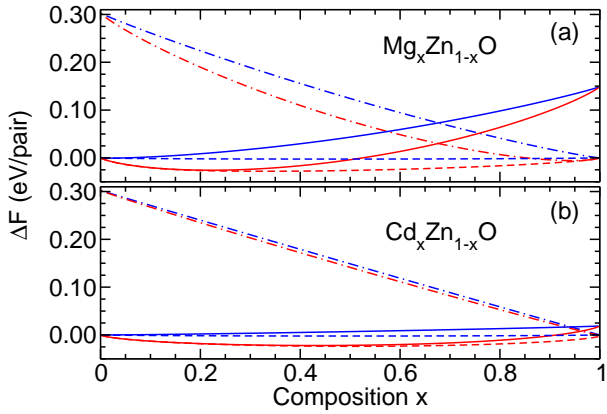}}
\caption{\label{fig:mixing_reference}(Color online) Mixing free energy $\Delta
  F(x,T)$ of Mg$_x$Zn$_{1-x}$O (a) and Cd$_x$Zn$_{1-x}$O (b) alloys versus
composition $x$. The curves are obtained using the GQCA with
wurtzite clusters only (solid), rocksalt clusters only (dash-dotted), or
both types of clusters (dashed). In all cases results are shown
for $T=300$~K (blue curves) and $T=1100$~K (red curves). The
equilibrium crystal structures $wz$ ($x=0$) and $rs$ ($x=1$) have been used as energy zero.}
\end{figure}

In the following, we compare to free energy curves that are computed
using the energies of the respective \emph{equilibrium} crystal
structures ($rs$-MgO, $wz$-ZnO, $rs$-CdO) as level of reference, in
contrast to the results in Fig.~\ref{fig:mixing} where the energies of
the binary components in the same crystal structure have been used.
For Mg$_x$Zn$_{1-x}$O the mixing free energy curves for the
$wz$ alloys and the $rs$ alloys in Fig.~\ref{fig:mixing_reference}(a)
intersect at $x\approx 0.67$, independent of the temperature. We
interpret this as a tendency for a transition from preferred fourfold
coordination ($wz$) to preferred sixfold coordination ($rs$) at that
composition under equilibrium conditions. In addition, the difference
of the mixing free energy per cluster of the heterostructural alloys
and that of the respective isostructural cases exceeds $25$~meV,
i.e.\ $k_\text{B}T$ at room temperature, for approximately $0.10 \leq x
\leq 0.98$ ($0.28 \leq x \leq 0.93$) at $T=300$~K ($T=1100$~K). The
additional degree of freedom of varying also the atomic coordination
leads to the lowest mixing free energy [see
Fig.~\ref{fig:mixing_reference}(a)]. Hence, for these values of $x$
both, $rs$ as well as $wz$ clusters, significantly contribute to the
alloy material.

The same tendencies have been observed in experimental studies.
Minemoto \emph{et al.}~\cite{Minemoto:2000} (Vashaei \emph{et
  al.}~\cite{Vashaei:2005}) found predominantly $wz$ structure
below $x\leq 0.46$ ($x\leq 0.34$) and mainly $rs$ structure for $x\geq 
0.62$ ($0.65 \leq x \leq 0.97$). Ohtomo
\emph{et al.}~\cite{Ohtomo:1998} report on the occurrence of an impurity
phase above Mg concentrations of $x\approx 0.33$ which roughly
coincides with the point in Fig.~\ref{fig:mixing_reference}(a) where the
high-temperature curve for $wz$ structure starts to deviate from the
curve for the mixed statistics. Also the X-ray diffraction
measurements of thin-film samples by Bundesmann \emph{et
  al.}~\cite{Bundesmann:2006} reveal hexagonal $wz$
structure for $x\leq 0.53$ and cubic $rs$ structure for $x\geq
0.67$. Films grown by reactive electron beam
evaporation\cite{Chen:2003} lead to hexagonal Mg$_x$Zn$_{1-x}$O up to
$x=0.51$ and to cubic 
Mg$_x$Zn$_{1-x}$O above $x=0.55$.
In general, the amount of fourfold- or sixfold-coordinated atoms
cannot be directly derived from the mixing free energy and will be
discussed in more detail in Section~\ref{sec:phase_transition}.

In the case of the Cd$_x$Zn$_{1-x}$O alloys, we study the mixing free
energy in Fig.~\ref{fig:mixing_reference}(b) where we observe a
crossing of the curves for the isostructural alloys roughly at a Cd
content of $x\approx 0.95$. Moreover, we find that the result from 
the mixed statistics differs less than $25$~meV (per cluster) from the
curve for the pure $wz$ structure up to compositions $x$ of about $0.17$
($0.59$) for $T=300$~K ($T=1100$~K). The reason for that behavior is
the small energy difference between the $rs$-CdO and the $wz$-CdO
phase~\cite{Schleife:2006} (see Tables~\ref{tab:params_wz}
and~\ref{tab:params_rs}).

Experimental studies of Cd$_x$Zn$_{1-x}$O show an ambivalent picture:
While two groups report very low thermodynamic solubility
limits~\cite{makino:2001} of only $x\approx 0.07$ or phase separation
at even lower Cd concentrations~\cite{bertram:2006}, another group
observed Cd concentrations up to $x=0.32$ in samples produced by means
of highly non-equilibrium MBE~\cite{Sadofev:2006}. Unfortunately, they
have not tried for higher concentrations. More importantly, the $wz$
crystal structure has also been found for plasma-enhanced MOCVD
layers~\cite{Shigemori:2004} up to $x=0.697$. Their result is
confirmed by Ishihara \emph{et al.}~\cite{Ishihara:2006} who report
a transition from $wz$ to $rs$ structure at $x=0.7$ for films 
deposited by MOCVD. Apparently, the change of the crystal structure
occurs at lower Cd concentrations than we predicted from the
intersection of the $\Delta F$ curves for $wz$ and $rs$ in
Fig.~\ref{fig:mixing_reference}(b). On the other hand, the
high-temperature curve from the mixed statistics shows significant
deviations from the pure $wz$ statistics above Cd contents of about 0.7
which may explain the experimental findings~\cite{Ishihara:2006}.

\begin{figure}
\resizebox{0.92\columnwidth}{!}{\includegraphics*{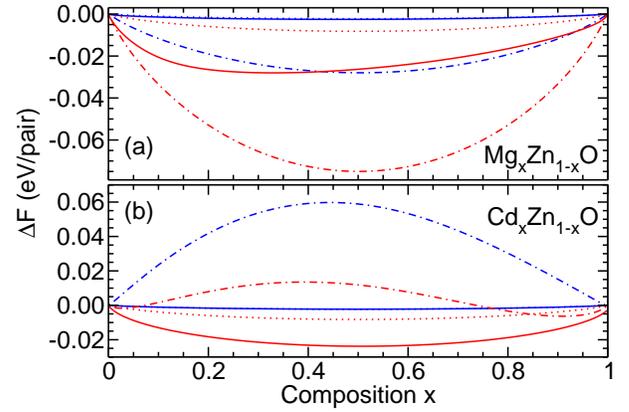}}
\caption{\label{fig:mixing_mix}(Color online) Mixing free energy $\Delta F(x,T)$ of
Mg$_x$Zn$_{1-x}$O (a) and Cd$_x$Zn$_{1-x}$O (b) alloys versus
composition $x$ for $T=300$~K (blue) and $T=1100$~K (red). The solid
curves are computed using cluster fractions from the GQCA. The dotted
curves are obtained for the MDM while the dash-dotted
curves are calculated using the ideal cluster fractions. All curves
result from the combined statistics with both wurtzite- and rocksalt-type
clusters. The respective composition end points have been used as
zero (see text).}
\end{figure}

For a detailed discussion of the mixed statistics we present in
Fig.~\ref{fig:mixing_mix} the results for the mixing free energy
obtained from the statistics with $wz$ and $rs$ clusters. In this plot
we use again the equilibrium crystal structures for each material as
level of reference for the mixing free energies at $x=0$ and
$x=1$. Since the ideal cluster fractions $x_j^0$ according to
Eq.~\eqref{eq:x_j^0} do not depend on the cluster energies, both the
sum of  the $x_j^0$ for all $wz$ clusters and also of all $rs$
clusters give the same total weight of 0.5. Consequently, at $x=0$ and
$x=1$ clusters of both crystal structures contribute
equally, albeit, we want to take this situation into account due
to possible non-equilibrium growth conditions. Therefore, we set the
mixing free energies at $x=0$ and $x=1$ to zero for each curve
resulting from the SRS model in
Fig.~\ref{fig:mixing_mix} since otherwise, $\Delta F(x=0,T)>0$ and $\Delta
F(x=1,T)>0$ hold for temperatures $T>0$~K due to the weights of the
SRS model.

First of all, Fig.~\ref{fig:mixing_mix} points out that the GQCA and
the MDM coincide for both materials at low temperatures. Contrary, the
SRS model and the GQCA result in a remarkable difference
independent of the temperature. This is not surprising, since the
SRS model neglects the large energetic differences
between the two crystal structures by merely assigning the ideal
weights to the clusters. While we find for Mg$_x$Zn$_{1-x}$O, which
shows negative excess energies, that the ideal cluster fractions are
energetically favored over the ones resulting from GQCA or the
MDM [cf.\ Fig.~\ref{fig:mixing_mix}(a)], the opposite is true for
Cd$_x$Zn$_{1-x}$O [cf.\ Fig.~\ref{fig:mixing_mix}(b)], where the excess
energies are positive.

\subsection{\label{sec:phase_transition}Phase transitions and phase diagram}

\begin{figure}
\resizebox{0.92\columnwidth}{!}{\includegraphics*{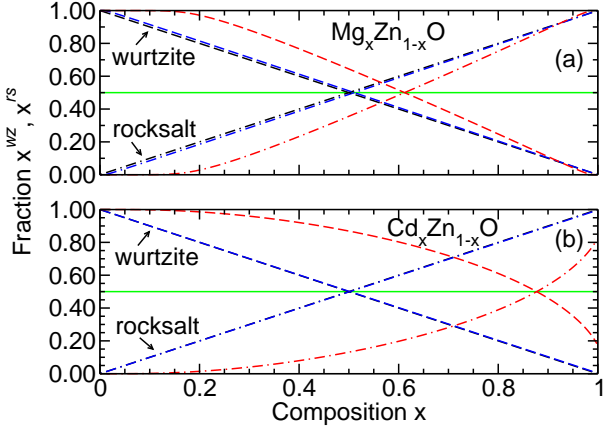}}
\caption{\label{fig:fractions}(Color online) Crystal-structure fractions $x^{wz}$
  (dashed curves) and $x^{rs}$ (dash-dotted curves) of
  Mg$_x$Zn$_{1-x}$O (a) and Cd$_x$Zn$_{1-x}$O (b) alloys versus
  composition $x$. We calculated the curves using the cluster
  fractions from the GQCA for $T=300$~K (blue curves) and $T=1100$~K
  (red curves).  The solid green lines are obtained using the ideal
  cluster fractions. For comparison the black curves show the
  result from the MDM for both crystal structures.}
\end{figure}

For a better understanding of the structural composition of the alloys
as a function of $x$ and $T$ we calculate the contributions of $wz$
and $rs$ clusters to the mixed statistics results. For that purpose we
define the $wz$ character $x^{wz}$ and the $rs$ character 
$x^{rs}$ of the alloy as the sum over the corresponding weights, i.e.,
\begin{equation}
\label{eq:mix_stat}
x^{wz}=\sum\limits_{j=0}^{21}{x_j}\text{ and } x^{rs}=\sum\limits^{37}_{j=22}x_j,
\end{equation}
with $x^{wz}+x^{rs}=1$. For given composition $x$, temperature $T$,
and statistical model $x^{wz}$ ($x^{rs}$) measures the relative
contribution of clusters with fourfold (sixfold) atomic coordination
to the studied (non-decomposed) X$_x$Zn$_{1-x}$O sample (X\,=\,Mg,Cd). The
results for $x^{wz}$ and $x^{rs}$ as a function of $x$ and $T$ are
shown for both pseudobinary materials in
Fig.~\ref{fig:fractions}. This figure clarifies how the dominating
crystal structure in the alloy depends on the thermodynamic 
conditions as well as the respective cluster statistics (GQCA, SRS, MDM). As discussed before, the SRS model for the
cluster fractions gives rise to equal contributions of $rs$ and $wz$
clusters (see Fig.~\ref{fig:fractions}). 
We confirm our discussion of the results for the mixing free energy by
finding the GQCA curves close to the MDM results for both
materials at room temperature. As expected, higher
preparation temperatures tend to move the intersection $x^{wz}=x^{rs}$
to larger Mg or Cd molar fractions $x$. More specifically, we find that
point at about $x=0.5$ ($T=300$~K) and $x=0.6$ ($T=1100$~K) for
Mg$_x$Zn$_{1-x}$O, while the temperature dependence is more pronounced
for Cd$_x$Zn$_{1-x}$O where the intersection for $T=1100$~K occurs at
about $x\approx 0.87$. Consequently, the local crystal structure and bonding
configuration of the Cd$_x$Zn$_{1-x}$O alloy depends much more on the
actual growing conditions which explains the ambivalent experimental
findings for that material
system.\cite{makino:2001,bertram:2006,Sadofev:2006,Shigemori:2004,Ishihara:2006}

\begin{figure}
\resizebox{0.92\columnwidth}{!}{\includegraphics*{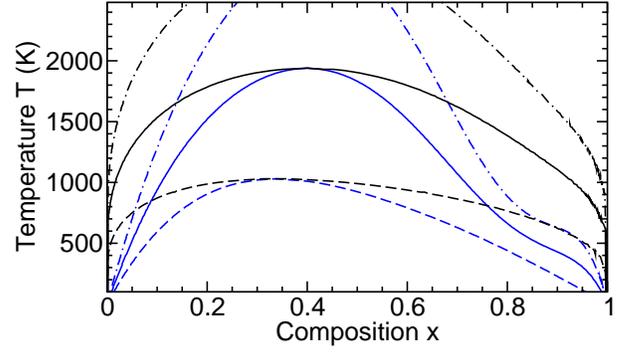}}
\caption{\label{fig:phase_diagram}(Color online) Phase diagram computed for Cd$_{x}$Zn$_{1-x}$O
  using the SRS model for the cluster statistics. We show binodal (black) and spinodal (blue) curves for
  $wz$ clusters only (dashed curves), $rs$ clusters only (dash-dotted
  curves), as well as for the mixed statistics including both sixfold
  and fourfold atomic coordination (solid curves).}
\end{figure}

As discussed before, the occurrence of two pronounced minima as well as
inflection points in the curve for the mixing free energy of
Cd$_x$Zn$_{1-x}$O obtained from the SRS model
(cf.\ Fig.~\ref{fig:mixing_mix}) is an indication for a possible phase
separation.\cite{Chen:1995} Due to its negative excess energies
[cf.\ Fig.~\ref{fig:excess}(a)] the Mg$_x$Zn$_{1-x}$O alloy
does not show such a behavior. To study the consequences we construct the
common tangent line to the free energy curves obtained from the
SRS model for Cd$_x$Zn$_{1-x}$O alloys at several
$T$. The resulting tangent points $x_1=x_1(T)$ and $x_2=x_2(T)$ describe the
boundaries of the miscibility gap and their variation with the
temperature defines the binodal line in the $T$\,-\,$x$ phase diagram
shown in Fig.~\ref{fig:phase_diagram}. From the maximum of the
resulting $T$\,-\,$x$ curves we find the critical parameters $T_{\rm
  crit}=1030$~K and $x_{\rm crit}=0.34$ for the $wz$ statistics as
well as $T_{\rm crit}=1940$~K and $x_{\rm crit}=0.4$ for the mixed
statistics. For the pure $rs$ statistics the critical temperature is
larger than 2500~K.
Inside the miscibility gap the alloy consists of a mixture of
Cd$_{x_{1}}$Zn$_{1-x_{1}}$O and Cd$_{x_{2}}$Zn$_{1-x_{2}}$O with the
mixing free energy
\begin{equation}
\Delta F_{\rm mix}(x,T)=\frac{x_2-x}{x_2-x_1}\Delta F(x_1,T)+
\frac{x-x_1}{x_2-x_1}\Delta F(x_2,T),
\end{equation}
which is lower than $\Delta F(x,T)$.\\
Furthermore, the two inflection points $x_1'=x_1'(T)$ and
$x_2'=x_2'(T)$ of the $\Delta F(x,T)$ curve from the SRS
model for Cd$_x$Zn$_{1-x}$O [cf.\ Fig.~\ref{fig:mixing_mix}(b)] define
the spinodal curve in the phase diagram in
Fig.~\ref{fig:phase_diagram}. Inside the intervals $x_1<x<x_1'$ and
$x_2'<x<x_2$ the alloy is metastable against local decomposition due
to an energy barrier. Contrary, spontaneous decomposition into two
alloys with compositions $x_1$ and $x_2$ happens within the interval
$x_1'<x<x_2'$.\cite{Chen:1995}
The actual
distribution of the clusters depends via $x_1(T)$ and
$x_2(T)$ on the composition $x$ and temperature $T$
(cf.\ Fig.~\ref{fig:phase_diagram}).\\
The phase diagram in Fig.~\ref{fig:phase_diagram} suggests the
speculation that the low solubilities of $rs$-CdO in
$wz$-ZnO~\cite{makino:2001,bertram:2006} can be explained within the
SRS model. For alloy preparation conditions that lead to a
stochastic occupation of the cation lattice sites,
Fig.~\ref{fig:phase_diagram} confirms a large miscibility gap starting
at $x\approx 0.08$ at $T\approx 830$~K ($wz$ clusters only), $T\approx
1450$~K ($rs$ and $wz$ clusters), or $T\approx 2070$~K ($rs$ clusters
only).

\subsection{Cohesive energies}

\begin{figure}
\resizebox{0.92\columnwidth}{!}{\includegraphics*{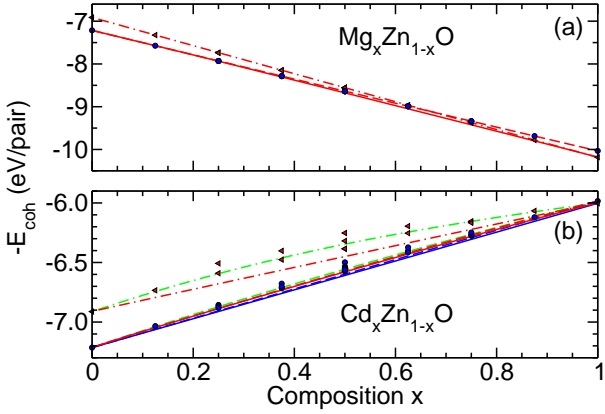}}
\caption{\label{fig:cohesion}(Color online) Negative cohesive energy
  $E_{\rm coh}$ (in eV per pair) of Mg$_x$Zn$_{1-x}$O (a) and
  Cd$_x$Zn$_{1-x}$O (b) alloys versus composition $x$. The curves are
  obtained using only wurtzite clusters (dashed line), only rocksalt
  clusters (dash-dotted line), or both type of clusters (solid
  line). We use cluster fractions from the GQCA for $T=300$~K (blue
  curves) and $T=1100$~K (red curves), as well as ideal cluster
  fractions (green curves). The MDM result coincides with the one for 
  from GQCA at room temperature. The cohesive energies are also shown
  for the clusters in rocksalt (red triangles) and wurtzite (blue circles) crystal
  structure.}
\end{figure}

From the cluster energies $\varepsilon_j$ as given in
Tables~\ref{tab:params_wz} and~\ref{tab:params_rs} we compute the
cohesive energies (per cation-anion pair) by subtracting the total
energies of the spin-polarized atoms, $\varepsilon_{\rm Mg}$, $\varepsilon_{\rm Zn}$,
$\varepsilon_{\rm Cd}$, and $\varepsilon_{\rm O}$, from the ground
state energy $\varepsilon_j$ of the $j$th cluster, i.e.,
\begin{equation}
E_{\text{coh},j} = -\varepsilon_j + \frac{n_j}{8}\varepsilon_{\text{X}} +
\frac{n-n_j}{8}\varepsilon_{\text{Zn}} + \varepsilon_{\text{O}},
\end{equation}
with X\,=\,Mg, Cd. For the binary end components this leads to
cohesive energies of $E_{\text{coh}}=10.18$~eV ($rs$-MgO), 7.21~eV
($wz$-ZnO), and 6.00~eV ($rs$-CdO) which are close to experimental
values of 10.26~eV, 7.52~eV, and 6.40~eV (see compilation in
Ref.~\onlinecite{Schleife:2006}). In Fig.~\ref{fig:cohesion} the
averaged cohesive energies [cf.\ Eq.~\eqref{eq:connolly}] per
cation-anion pair, as computed using the cluster weights from the different
statistics, are plotted for Mg$_{x}$Zn$_{1-x}$O and Cd$_{x}$Zn$_{1-x}$O
alloys versus composition $x$. From Fig.~\ref{fig:cohesion}(a) it immediately becomes
clear that for Mg$_{x}$Zn$_{1-x}$O the variation of the cohesive energy with
the alloy composition $x$ is almost linear. This is due to the fact
that the variation between the different crystal structures as well as the
variation of the excess energies with $x$ for fixed crystal structure
(cf.\ Fig.~\ref{fig:excess}) both are small compared to the large
difference of $E_\text{coh}$ between ZnO and MgO. For that reason also the temperature
variation of the GQCA results is not visible. Overall, neither for the
$rs$ nor the $wz$ crystal structure there is a visible dependence
on the actual statistics that has been used.

For Cd$_{x}$Zn$_{1-x}$O we find from Fig.~\ref{fig:cohesion}(b) that the
results within the GQCA and the SRS model differ. The
curves from the MDM cannot be distinguished from the low-temperature
GQCA results. For the $wz$ crystal structure deviations resulting from
the different statistics are small but visible. As discussed
before, the mixed statistics gives rise to results that are close to those
obtained when taking only $wz$ clusters into account.

\section{\label{sec_lattice}Structural and elastic properties}

\subsection{Fourfold bonding coordination}

\begin{figure}
\resizebox{0.92\columnwidth}{!}{\includegraphics*{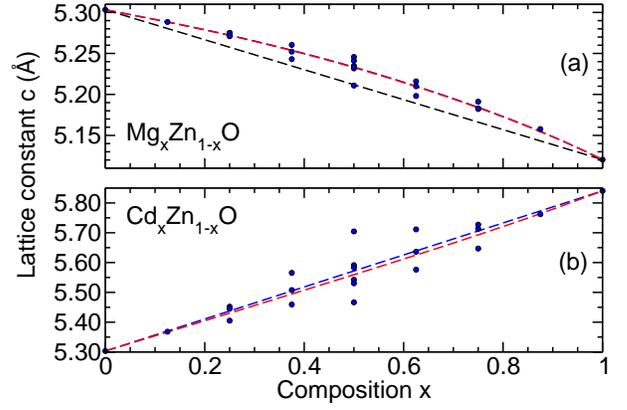}}
\caption{\label{fig:lattice_c}(Color online) Lattice parameter $c$ (in \AA) of the
  wurtzite structure for Mg$_x$Zn$_{1-x}$O (a) and Cd$_x$Zn$_{1-x}$O (b)
  alloys with clusters in $wz$ crystal structure only versus composition
  $x$. The black curves represent the MDM and the green curves the
  ideal cluster fractions. The GQCA results are shown for $T=300$~K
  (blue curve) and $T=1100$~K (red curve). For Mg$_x$Zn$_{1-x}$O the GQCA curves
  and the SRS results coincide. For Cd$_x$Zn$_{1-x}$O the
  low-temperature GQCA coincides with the MDM and the high-temperature
  GQCA with the result using ideal cluster fractions. Blue dots give
  the results for the individual clusters.}
\end{figure}

In Section~\ref{sec_thermo} we found that especially for the 
Cd$_x$Zn$_{1-x}$O system a significant fraction of the alloy locally
shows the $wz$ crystal structure over a large composition range. For
that reason we investigate the behavior of the $c$ lattice constant and
the $u$ parameter as characteristic parameters of the $wz$ structure.

Figure~\ref{fig:lattice_c}(a) shows that in the case of Mg$_x$Zn$_{1-x}$O
the results from the GQCA for both temperatures and the curve obtained
using the ideal cluster fractions almost coincide. The two statistics give rise to slightly larger values for $c$
than the MDM and, therefore, also Vegard's law, which we find to be
violated. Around $x=0.5$ the deviation may reach values of about
0.5\,\%. As can be seen in Fig.~\ref{fig:lattice_c}(a) this leads to
large errors when the composition of an alloy is determined by 
measuring the $c$-lattice constant and applying Vegard's rule.

Contrary, for Cd$_x$Zn$_{1-x}$O [cf.\ Fig.~\ref{fig:lattice_c}(b)] the
curves computed with the GQCA cluster fractions for high and low
temperatures represent the two limiting cases. While the
low-temperature curve coincides with the 
MDM, the high-temperature curve is close to the results from the
SRS model and the deviations between the two cases are
small.

A comparison of the values $c_j$ for the individual clusters of both
material systems shows that the variation is much
larger for Cd$_{n_j}$Zn$_{n-n_j}$O$_n$ than for
Mg$_{n_j}$Zn$_{n-n_j}$O$_n$. Overall, the deviations from the MDM
(which is equivalent to Vegard's law) are larger for Mg$_x$Zn$_{1-x}$O and we find a
remarkable bowing of the curves which renders a purely linear
interpolation questionable. Contrary, the bowing is smaller for
Cd$_x$Zn$_{1-x}$O despite the stronger deviations for individual
clusters, e.g., with $n_j=3$, $4$, or $5$ Cd atoms.

\begin{figure}
\resizebox{0.92\columnwidth}{!}{\includegraphics*{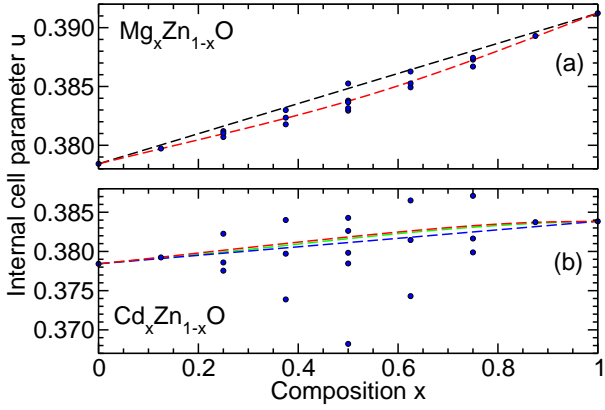}}
\caption{\label{fig:lattice_u}(Color online) Internal cell parameter $u$ of the wurtzite
  structure for Mg$_x$Zn$_{1-x}$O (a) and Cd$_x$Zn$_{1-x}$O (b) alloys
  derived from clusters in $wz$ crystal structure only versus composition $x$. The black
  curves represent the MDM and the green curves the SRS
 model. The GQCA results are shown for $T=300$~K (blue curve) and 
  $T=1100$~K (red curve). For Mg$_x$Zn$_{1-x}$O the GQCA curves and the ideal
  cluster fractions coincide. For Cd$_x$Zn$_{1-x}$O the low-temperature GQCA
  coincides with the MDM. Blue dots give the results for the individual
  clusters involved.}
\end{figure}

In Fig.~\ref{fig:lattice_u} we also study the influence of the cluster
statistics on the internal cell parameter $u$ of the hexagonal
lattice, which describes the translation of the cation sublattice with
respect to the anion sublattice along the $c$ direction. Again, for 
Mg$_x$Zn$_{1-x}$O there is a noticeable deviation of all statistics
(that are almost equal) from the MDM result, with a remarkable bowing of the
curves. The $u$ values of the individual clusters show a larger
scattering for Cd$_{n_j}$Zn$_{n-n_j}$O$_n$ than for
Mg$_{n_j}$Zn$_{n-n_j}$O$_n$. In the case of Cd$_x$Zn$_{1-x}$O again the
low-temperature GQCA curve coincides 
with the MDM result, while the high-temperature GQCA 
curve is close to the one obtained using ideal cluster fractions.

\subsection{Cation-cation distance}

\begin{figure}
\resizebox{0.92\columnwidth}{!}{\includegraphics*{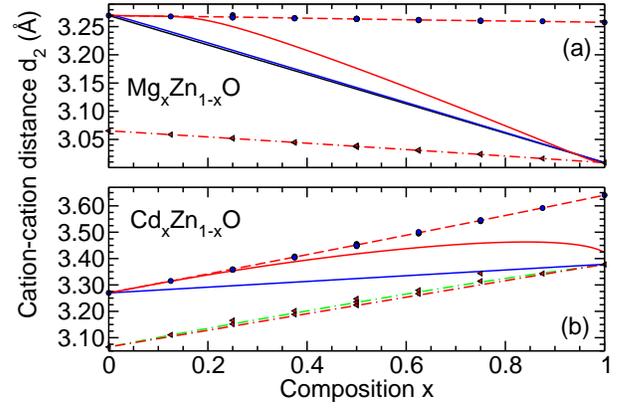}}
\caption{\label{fig:lattice_a0}(Color online) Average cation-cation (second-nearest
  neighbor) distance $d_2$ of Mg$_x$Zn$_{1-x}$O
  (a) and Cd$_x$Zn$_{1-x}$O (b) alloys versus composition $x$. The
  curves are obtained using the GQCA with wurtzite clusters only
  (dashed curves), rocksalt clusters only (dash-dotted curves), or
  both types of clusters (solid curves). For the GQCA we present
  results for $T=300$~K (blue curves) and $T=1100$~K (red curves). For
  comparison we show results obtained using the MDM (black
  curves) and the SRS model (green curves). The
  cation-cation distances are also depicted for the individual clusters in rocksalt
  (red triangles) and wurtzite (blue circles) crystal structure.}
\end{figure}

Both crystal structures that are taken into account to describe the
pseudobinary alloy system, $rs$ and $wz$, are characterized by
different lattice parameters. While the cubic $rs$ structure is
described by only one lattice constant $a_0$, there are three
independent parameters $a$, $c$, and $u$ for the hexagonal $wz$
structure which renders a universal description of the lattice structure
of the alloys more difficult. As mentioned before, we solve this
problem by restricting our discussion to the average cation-cation
distance $d_2(x)$. This quantity can be defined in a unique 
manner even for different crystal structures and is experimentally
accessible at the same time. For the $rs$ lattice $d_2$ is simply
related to the lattice constant $a_0$ by $d_2=a_0/\sqrt{2}$ and for
the ideal $wz$ lattice with $c/a=\sqrt{8/3}$ it holds $d_2=a$.
We average over all cation-cation distances in one cluster to obtain
the $d_{2,j}$ (cf.\ Tables~\ref{tab:params_wz}
and~\ref{tab:params_rs}). Using the respective weights from the GQCA,
the SRS model, and the MDM, the configurational average is calculated
from Eq.~\eqref{eq:connolly} and the results are plotted in
Fig.~\ref{fig:lattice_a0} for Mg$_x$Zn$_{1-x}$O and Cd$_x$Zn$_{1-x}$O.

For Mg$_x$Zn$_{1-x}$O [cf.\ Fig.~\ref{fig:lattice_a0}(a)] we find for the
pure $wz$ and pure $rs$ cluster geometries that the three alloy
statistics give almost the same results and, therefore, also agree
well with Vegard's rule. For each fixed crystal structure the
variation of $d_2$ versus $x$ is small, whereas $d_2$ differs up to
8\,\% between the $wz$ and the $rs$ structure of a material. While the
GQCA results from the mixed statistics agrees well with the MDM for
$T=300$~K, we find significant deviations for $T=1100$~K. In the
high-temperature case the cation-cation distance in the pseudobinary alloy remains very
close to the value of the pure $wz$ structure up to compositions of
about $x\approx 0.2$ [cf.\ Fig.~\ref{fig:lattice_a0}(a)]. For higher Mg
contents $d_2$ approaches the value of $rs$-MgO.

For Cd$_x$Zn$_{1-x}$O [cf.\ Fig.~\ref{fig:lattice_a0}(b)] the variation
of $d_2$ is much larger than for Mg$_x$Zn$_{1-x}$O since the
cation-cation distance varies much more between ZnO and CdO. The
results from all three statistics for the pure $wz$ clusters and the
pure $rs$ clusters agree well with Vegard's rule except of a small
deviation for the SRS model for $rs$-CdO. Again the most
remarkable change is observed for the mixed cluster expansion. At room
temperature the curve connects the $wz$ and $rs$ end
components almost linearly. For the higher temperature of $T=1100$~K the cation-cation
distance closely follows the $wz$ trend up to Cd concentrations of
more than $0.5$. Overall, the different structural configurations can
be clearly distinguished at fixed $x$ via the average cation-cation distance. When
both lattice structures can occur in the alloy we even find a more
pronounced temperature dependence.

\subsection{Bulk modulus}

\begin{figure}
\resizebox{0.92\columnwidth}{!}{\includegraphics*{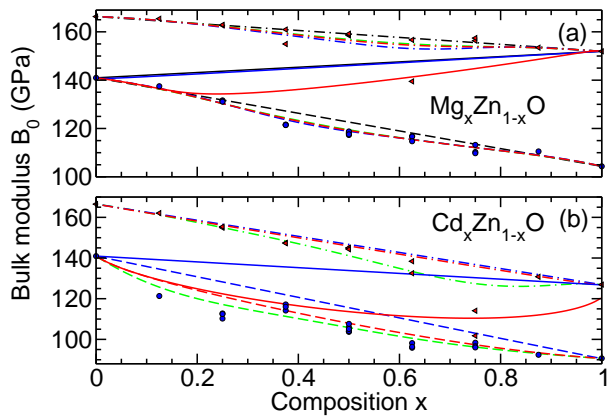}}
\caption{\label{fig:bulk}(Color online) Bulk modulus $B_0$ (in GPa) of
  Mg$_x$Zn$_{1-x}$O (a) and Cd$_x$Zn$_{1-x}$O (b) alloys versus
  composition $x$. The curves are obtained using the GQCA with 
  wurtzite clusters only (dashed curves), rocksalt clusters only
  (dash-dotted curves), or both type of clusters (solid curves). For
  the GQCA we present results for $T=300$~K (blue curves) and
  $T=1100$~K (red curves). For comparison we also show results
  obtained using the MDM (black curves) and the SRS model (green
  curves). The bulk moduli are also depicted for the individual
  clusters in rocksalt (red triangles) and wurtzite (blue circles) crystal structure.}
\end{figure}

Besides the cation-cation distance we also study the bulk modulus
$B_0$ for Mg$_x$Zn$_{1-x}$O and Cd$_x$Zn$_{1-x}$O using the three
statistics evolving from the cluster weights of the GQCA, the SRS
model, and the MDM. The results are plotted together with the
values for the individual clusters in Fig.~\ref{fig:bulk}.

For Mg$_x$Zn$_{1-x}$O [cf.\ Fig.~\ref{fig:bulk}(a)] in the pure $rs$ or
$wz$ structure the GQCA results show a temperature dependence and also
differ from the linear interpolation of the MDM, whereas the SRS curve
is very close to the high-temperature result of the 
GQCA. The reason for this behavior is the stronger deviation of the
bulk moduli for the individual clusters from the linear
interpolation. As can be seen in Fig.~\ref{fig:bulk}(a) for the $rs$
structure we find a certain cluster
with an especially low bulk modulus at $x=0.625$ which, therefore,
coincides with the minimum of the GQCA curves. For the GQCA results
using the mixed statistics we find a strong temperature dependence,
since the room-temperature curve closely follows the MDM curve, while
the high-temperature result matches the pure $wz$ curve up to
$x\approx 0.2$ and then approaches the pure $rs$ curve at $x=1$.

In the case of Cd$_x$Zn$_{1-x}$O [see Fig.~\ref{fig:bulk}(b)] we find both
GQCA curves for the pure $rs$ lattice structure close to the MDM
result. This is only true for the low-temperature result of the pure
$wz$ structure, whereas the high-temperature curve differs remarkably in
this case. For both $rs$ and $wz$ structure, the SRS model
shows the largest deviation from the linear interpolation. Again this
can be attributed to the stronger variation of the values for $B_0$ of
the individual clusters with respect to the MDM curve with the
strongest deviation occurring for $rs$ at $x=0.75$ [cf.\
Fig.~\ref{fig:bulk}(b)]. For the mixed statistics for Cd$_x$Zn$_{1-x}$O
we find again that the low-temperature curve closely follows the
linear interpolation, while the high-temperature curve shows pronounced
deviations. At low concentrations it follows the high-temperature
GQCA curve of the pure $wz$ structure and does not even fully
approach the $rs$ limit due to the very small energetical difference
of these two polymorphs for CdO (cf.\ Fig.~\ref{fig:fractions}).

\section{\label{sec_conclusion}Summary and conclusions}

In summary, we studied pseudobinary A$_x$B$_{1-x}$C alloys whose binary end
components crystallize in different crystal structures under ambient
conditions. To study Mg$_x$Zn$_{1-x}$O and Cd$_x$Zn$_{1-x}$O with
$rs$ (MgO, CdO) and $wz$ (ZnO) equilibrium lattice structures
we adopted the cluster expansion for $wz$ and presented a
corresponding expansion for 16-atom $rs$ clusters. To include
simultaneously locally different bonding configurations with
either fourfold ($wz$) or sixfold ($rs$) coordination the GQCA has
been extended correspondingly. Besides the GQCA which minimizes the
mixing free energy with respect to the cluster fractions $x_j$ we took
two other distributions $x_j$ into account to simulate also certain
non-equilibrium growth procedures. We have combined these methods 
with \emph{ab initio} density functional calculations using the
semi-local GGA to treat exchange and correlation.
This provides knowledge of structural and energetic properties for each of
the clusters which we use together with the cluster
fractions to perform the configurational averages and, hence, to derive
an averaged alloy property for given composition and temperature.

Correspondingly, we find that the alloys' thermodynamics is determined 
by the local chemical bonding and the cluster statistics and,
therefore, the conditions under which the alloy is
prepared. The mixing free energies indicate a change from preferred
$wz$ crystal structure to preferred $rs$ crystal structure at about
$x\approx 0.675$ for Mg$_x$Zn$_{1-x}$O and about $x\approx 0.95$ for
Cd$_x$Zn$_{1-x}$O. We find that for Mg$_x$Zn$_{1-x}$O as well as
Cd$_x$Zn$_{1-x}$O the random alloy is always the most favorable. Only
for Cd$_x$Zn$_{1-x}$O in the SRS limit we observe the
occurrence of a miscibility gap and binodal as well as spinodal
decomposition ranges with the critical
parameters $T_{\rm crit}=1030$~K and $x_{\rm crit}=0.34$ for pure $wz$
structure, $T_{\rm crit}=1940$~K and $x_{\rm crit}=0.4$ for the mixed
statistics, and a critical temperature above 2500~K for pure $rs$
crystal structure. In any case we were able to calculate even the
temperature- and composition-dependent amount of $wz$ and $rs$
clusters in the system. The resulting ambivalent picture which is
strongly dependent on the actual conditions, is verified by a variety
of experimental results. For both material systems there are
experiments that agree with our predictions but also other experiments 
with deviating findings, showing that also measured properties of alloy
films have to be related to growth conditions, deposition method,
possible subsequent temper steps, etc.

For structural properties such as cation-cation distances and bulk
moduli we also find a significant dependence on the statistics that has
been employed for the alloy modeling. In addition, the pronounced
temperature dependence of the structural parameters and also the
boundary conditions due to the growth mechanism should lead to
distinct experimental observations of the alloy composition.

\section*{Acknowledgments}

We thank Lara K.\ Teles and R.\ Goldhahn for valuable
discussions. The research leading to these results has received
funding from the European Community's Seventh Framework
Programme (FP7/2007-2013) under grant agreement No.\ 211956 and by the
Deutsche Forschungsgemeinschaft (Project No.\ Be
1346/20-1). We acknowledge grants of computer time from the
Supercomputerzentrum Stuttgart and the HPC-EUROPA2 project (project number:
228398) with the support of the European Commission\,--\,Capacities
Area\,--\,Research Infrastructures. A.\ Schleife thanks the
Carl-Zeiss-Stiftung for support.

\end{document}